\documentclass[manuscript]{acmart}
\usepackage{graphicx} 
\usepackage{svg} 
\usepackage{xcolor}
\usepackage{soul}
\usepackage{lipsum}
\usepackage{glossaries}
\usepackage{lscape}
\usepackage{pdflscape}

\usepackage{adjustbox}
\usepackage{array}
\newcolumntype{R}[2]{%
    >{\adjustbox{angle=#1,lap=\width-(#2)}\bgroup}%
    l%
    <{\egroup}%
}
\usepackage{makecell}

\newcommand*\rot{\multicolumn{1}{R{45}{1em}}}

\usepackage{wasysym}
\newcommand{\ok}{\CIRCLE}
\newcommand{\nok}{\Circle}
\newcommand{\hok}{\LEFTcircle}

\usepackage{pifont}
\newcommand{\cmark}{\ding{51}}
\newcommand{\xmark}{\ding{55}}

\usepackage[most]{tcolorbox}
\newtcolorbox{Takeaway}{
  colback=gray!10,
  colframe=black!75,
  coltitle=black,
  fonttitle=\bfseries,
  title=Takeaway,
  attach boxed title to top center={yshift=-2mm,xshift=-60mm},
  boxed title style={
    colback=gray!60,
    colframe=black,
    boxrule=0.8pt,
    arc=1mm,
  },
  arc=1mm,
  drop shadow,
  enhanced,
  width=\linewidth
}

\title[Abstraction of Trusted Execution Environments as the Missing Layer for Broad Confidential Computing Adoption]{Abstraction of Trusted Execution Environments as the Missing Layer for Broad Confidential Computing Adoption: A Systematization of Knowledge}

\author{Quentin Michaud}
\email{quentin.michaud@telecom-sudparis.eu}
\affiliation{
    \institution{SAMOVAR, Télécom SudParis, Institut Polytechnique de Paris \& Thales, cortAIx Labs}
    \country{Palaiseau, France}
}
\author{Sara Ramezanian}
\email{sara.ramezanian@kau.se}
\affiliation{
    \institution{Lund University \& Karlstad University}
    \country{Lund, Sweden, \& Karlstad, Sweden}
}
\author{Dhouha Ayed}
\email{dhouha.ayed@thalesgroup.com}
\affiliation{
    \institution{Thales, cortAIx Labs}
    \country{Palaiseau, France}
}
\author{Olivier Levillain}
\email{olivier.levillain@telecom-sudparis.eu}
\affiliation{
    \institution{SAMOVAR, Télécom SudParis, Institut Polytechnique de Paris}
    \country{Palaiseau, France}
}
\author{Joaquin Garcia-Alfaro}
\email{joaquin.garcia_alfaro@telecom-sudparis.eu}
\affiliation{
    \institution{SAMOVAR, Télécom SudParis, Institut Polytechnique de Paris}
    \country{Palaiseau, France}
}

\newacronym{cc}{CC}{Confidential Computing}
\newacronym{ccc}{CCC}{Confidential Computing Consortium}
\newacronym{tee}{TEE}{Trusted Execution Environment}
\newacronym{tcb}{TCB}{Trusted Computing Base}
\newacronym{vm}{VM}{Virtual Machine}
\newacronym{os}{OS}{Operating System}

\begin{document}

\begin{abstract}
Trusted Execution Environments (TEEs) protect sensitive code and data from the operating system, hypervisor, or other untrusted software. Different solutions exist, each proposing different features. Abstraction layers aim to unify the ecosystem, allowing application developers and system administrators to leverage confidential computing as broadly and efficiently as possible. We start with an overview of representative available TEE technologies. We describe and summarize each TEE ecosystem, classifying them in different categories depending on their main design choices. Then, we propose a systematization of knowledge focusing on different abstraction layers around each design choice. We describe the underlying technologies of each design, as well as the inner workings and features of each abstraction layer. Our study reveals opportunities for improving existing abstraction layer solutions. It also highlights WebAssembly, a promising approach that supports the largest set of features. We close with a discussion on future directions for research, such as how future abstraction layers may evolve and integrate with the confidential computing ecosystem.
\end{abstract}

\maketitle

\section{Introduction}

In the last decades, information technologies have seen a large increase in usage, and so has data collection, including private or sensitive information. Furthermore, the size and complexity of these systems makes difficult their management, which is often delegated to third parties. This makes the critical data of an entity (be it an individual, a company, or a government) accessible to these third parties, which are often untrusted.

To tackle this problem, various security mechanisms have successfully been designed to protect data at rest~\cite{ieee_encrypted_storage} and in transit~\cite{rfc8446}.
However, as computing shifts towards public clouds, the ecosystem severely lacks a way to execute arbitrary computations on data while protecting said data against the untrustworthy platform that is providing the computing resources.
Confidential computing is a technology that aims to tackle these problems, by executing computations in a secure, hardware-based, attested space named a \gls{tee}.

A TEE's reliance on hardware fractured the confidential computing ecosystem across hardware vendors who each proposed their own incompatible, in-house solution.
Furthermore, most confidential computing solutions require the adaptation of preexisting software. This makes the development of generic, broadly available confidential-computing applications complex.
It also forces the developers to focus on the specifics of confidential computing.
This can also incur security risks if an application leveraging confidential computing does so in an unsafe way, possibly defeating the purpose of confidential computing.

In order to simplify the work of such developers, several technologies building upon confidential computing emerged. These technologies have various goals, ranging from allowing developers to target different confidential computing solutions at once to supporting common software distribution formats such as containers.

\subsection{Motivation}

In the age of a growing need for data security, confidential computing became a critical area of research and development.
However, despite its potential, the adoption of confidential computing technologies remains hindered by complexities related to their implementation and integration.
The main issue comes from the disparity of available hardware and technologies.
In a survey on confidential computing research, Feng et al.~\cite{feng_surveycc_2024} concluded that \textit{the absence of standardized implementation methods has resulted in diverse approaches by different vendors. This variance introduces technical complexity, bringing challenges even for professional security staff.}

At the core of this challenge lies the need for additional technologies, that we may call abstraction layers, that simplify the interaction with underlying hardware and security mechanisms, making it easier for developers to deploy secure solutions without deep expertise in the low-level details of the underlying confidential computing solutions.
These abstraction layers have the potential to bridge the gap between complex security infrastructure and practical applications, promoting broader adoption of confidential computing in a landscape of situations ranging from cloud computing to IoT.

This systematization of knowledge aims to explore and categorize the existing abstraction layers in confidential computing, offering insights into their strengths, weaknesses, and areas for future innovation.
By identifying key trends, frameworks, and challenges, this study lays the foundation for a deeper understanding of how abstraction layers can evolve to make confidential computing more accessible and efficient.

\subsection{Scope}

As abstraction layers are mainly relevant in the way the confidential workloads are executed, this study is focusing on \gls{tee}s, ignoring other aspects of confidential computing such as attestation. 
More precisely, this study is focusing on TEEs implemented on Application-Specific Integrated Circuits (ASICs) that support multiple CPU modes (and as such are able to run OSes relying on preemptive multitasking) and excludes technologies that provide similar security guarantees in other hardware domains, such as GPU TEEs~\cite{graviton,strongbox}, FPGA-based enclaves~\cite{shef_fpga_enclaves,fpga_enclaves}, or microcontrollers TEEs~\cite{noorman2013sancus,noorman2017sancus2,esptee}.
While these technologies offer valuable security features, their architectures, threat models, and use cases differ significantly from CPU TEEs, making direct comparisons less meaningful.
Additionally, GPU TEEs and other non-CPU secure enclaves are still in early research or limited to specialized applications, whereas CPU TEEs are widely deployed in commercial processors and cloud environments, where having the possibility to develop one application that targets multiple technologies is relevant.

\subsection{Contributions}

Several surveys tackled the confidential computing ecosystem~\cite{mo_machine_2024,sardar_confidential_2023,wang_confidential_2024,li2024sokpitfallstee,feng_surveycc_2024}. However, they mainly focus on the low-level details of the TEE solutions, and they often do not make a clear distinction between TEE technologies and software leveraging these TEEs. Furthermore, they fail to rigorously list and analyze existing abstraction layers, along with considering ways for improving them. This survey focuses on abstraction layers, but keep a discussion on TEEs to introduce concepts needed for discussing abstraction layers.

The main contributions of this survey are the following:
\begin{itemize}
    \item A detailed overview of the \gls{tee} ecosystem, focused on how these solutions are designed. This paper gives essential insights in the security challenges of the different approaches.
    \item An in-depth systematization of knowledge on existing abstraction layers technologies and their uses, its approaches and limitations.
    \item A list of open issues and future challenges regarding the development and the deployment of portable applications protected by confidential computing.
\end{itemize}





\subsection{Paper organization}

\begin{figure}[!hptb]
    \centering
    \includegraphics[width=\linewidth]{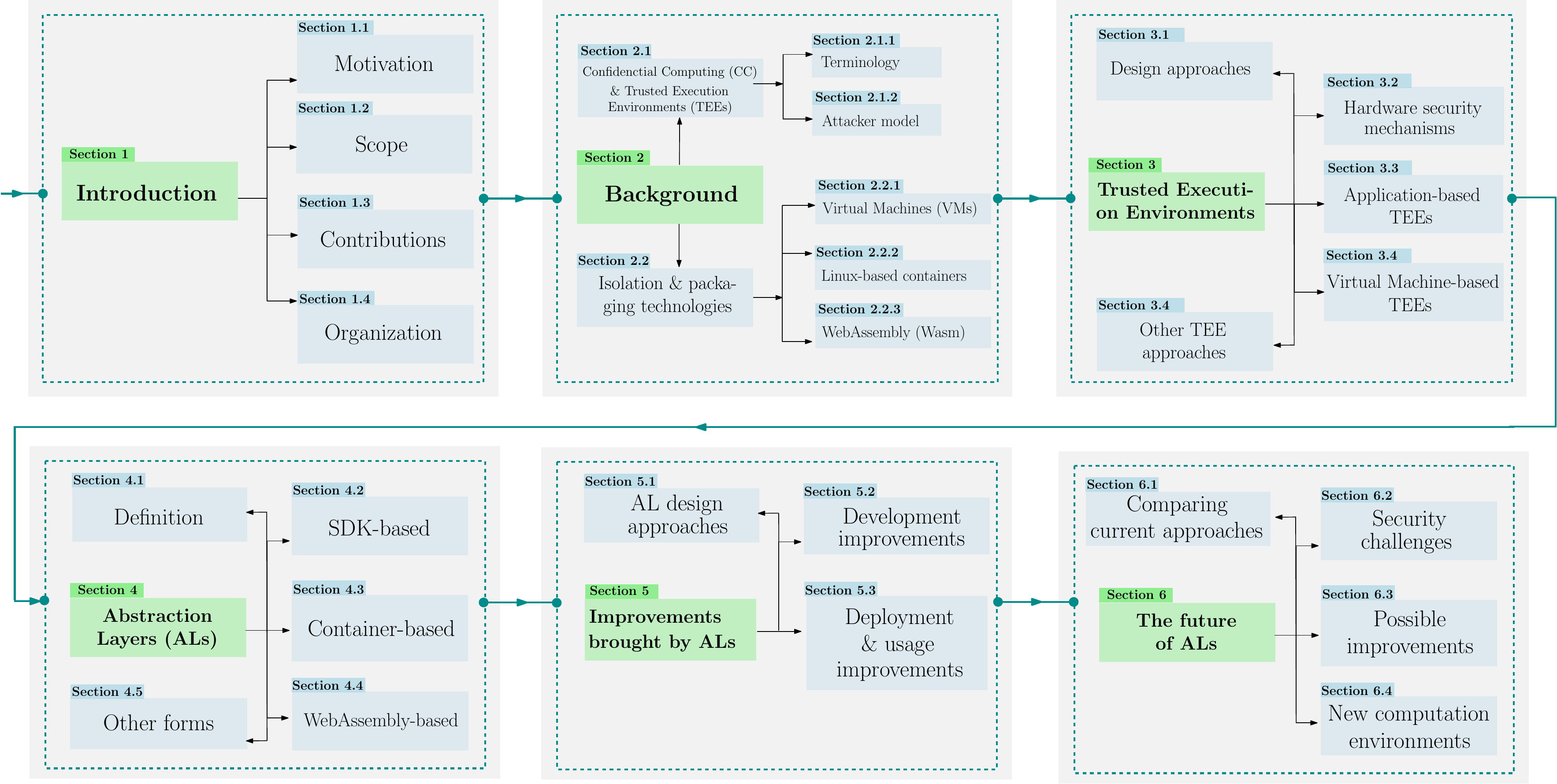}
    \caption{Organization of the main sections.}
    \label{fig:roadmap}    
\end{figure}

The organization of this paper is summarized in Figure~\ref{fig:roadmap}. Section~\ref{sec:background} provides necessary background on confidential computing and technologies related to the concept of abstraction layers. Then, Section~\ref{sec:tee} presents the main TEE technologies existing in the ecosystem, with a quick overview of how they work. Section~\ref{sec:al} then describes the abstraction layers, existing approaches for developing them, and their inner workings. Discussion presented in Section~\ref{sec:discussion} explores the ups and downs of abstraction layers, especially regarding the approach used to develop then. Finally, we discuss the future of confidential computing abstraction layers in Section~\ref{sec:future} before concluding in Section~\ref{sec:conclusion}.

\section{Background}
\label{sec:background}

This section presents the necessary background for this systematization of knowledge. We first present the confidential computing ecosystem, focusing on the definition of TEE, before presenting various isolation and packaging technologies that are often leveraged by abstraction layers. 

\subsection{Confidential Computing and Trusted Execution Environments}

Confidential computing is a generic term regrouping several mechanisms that we define and explain in the next sections.

\subsubsection{Terminology}

The \gls{ccc} defines confidential computing as \textit{the protection of data-in-use by performing computation in a hardware-based, attested Trusted Execution Environment (TEE)}~\cite{ccc2022technicalanalysis}. The \gls{ccc} further defines a TEE as \textit{an environment that provides a level of assurance regarding three properties: data confidentiality, data integrity and code integrity}. Confidential computing includes not only the environment in which the computation will take place (the TEE), but also other aspects such as attestation.

In previous work~\cite{sardar_confidential_2023}, the \gls{ccc} definitions have been criticized as being imprecise. One of the issues is that the \gls{ccc} distinguishes between hardware-based TEEs and virtualized, software TEEs, but no formal definition is given for these categories. Since the definition of confidential computing includes hardware-based TEE but not software TEEs, a lack of definition prevents an accurate listing of TEEs that can be included in the definition of confidential computing. Furthermore, it is unclear if "hardware-based" includes firmware, which can be considered a form of software.

For these reasons, this paper proposes a refined interpretation of the \gls{ccc} definition of a TEE, that will be used throughout the paper. The goal is to make this definition as broad as possible, so that it encompasses the \gls{ccc} definition as well as other approaches that may not fit the latter but remain relevant within the ecosystem.
This definition is as follows: a hardware-based TEE is an environment that provides a level of assurance regarding data confidentiality, data integrity and code integrity, and which relies on assistance from the hardware to this aim. This means that we consider TEEs that are mainly software-based, but leverage hardware memory protections.
In the rest of the paper, we use the term \emph{enclave} to refer to a concrete instantiation of a TEE-protected application. While this term originates from Intel SGX, it is now broadly used, even for applications instantiated within other TEE technologies. The paper will prefer using this term generically instead of other TEE-specific terms.

\subsubsection{Attacker model}
\label{sec:attacker-model}

Confidential computing considers a strong adversary that has access to the complete system software stack, including bootloaders, operating systems, firmware, and hypervisors. From these locations, the adversary can carry out all kind of attacks against the TEE that may compromise the confidentiality of the data inside the TEE.
However, a TEE cannot protect the data it contains if the code running inside the TEE is vulnerable. Attacks gaining access to the TEE content by exploiting the application running inside the TEE are out of scope.

Hardware attacks are also out-of-scope. While some TEE technologies may provide additional protection against hardware attacks (e.g. by encrypting memory contents), this attack vector will not be considered as an evaluation factor in this study.
Finally, a TEE cannot protect against Denial-of-Service (DoS) attacks, since it is impossible to prevent an adversary having access to low-level software stacks, such as the OS, from stopping the TEE by simply stopping the machine.

\subsection{Isolation and packaging technologies}

\begin{figure}[!hptb]
    \centering
    \includegraphics[width=\linewidth]{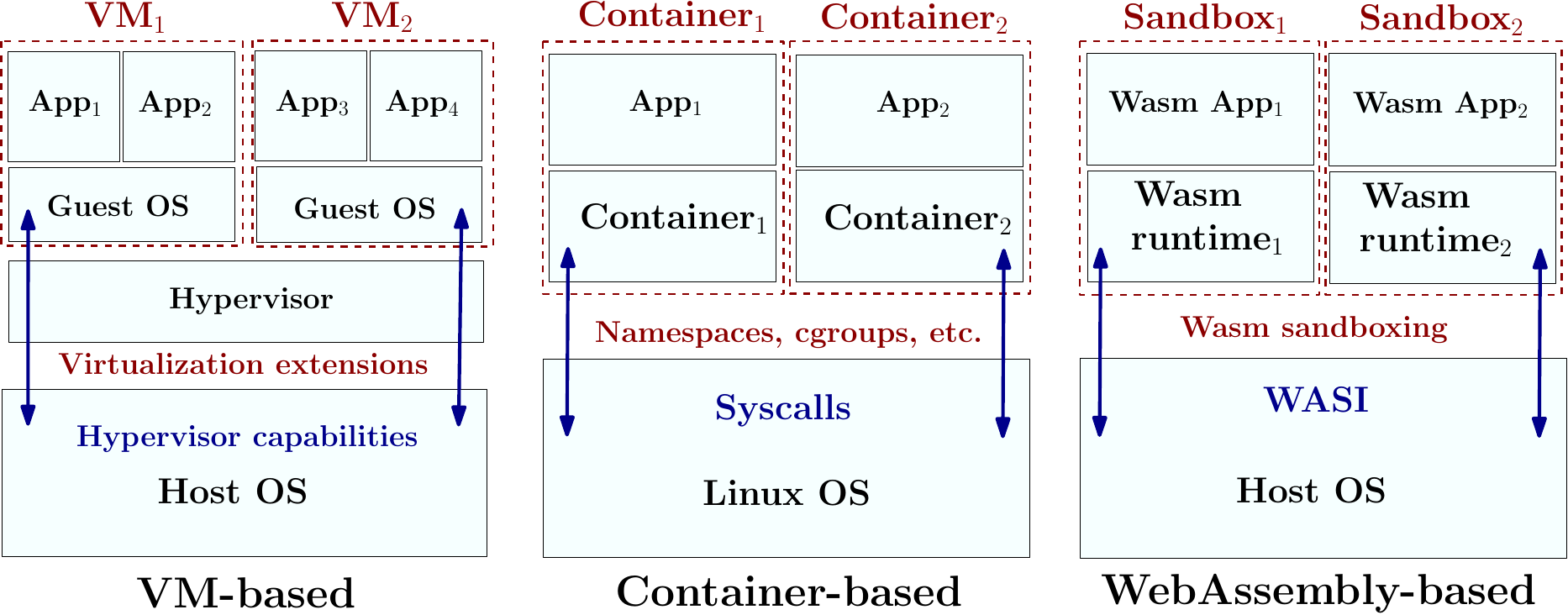}
    \caption{An overview of each isolation technology and their interactions with the untrusted host. The red dashed color represents each isolation technology. The blue color represents the interfaces that interact outside the isolation solution.
    }
    \label{fig:isolation_overview}
\end{figure}

Virtual machines, Linux-based containers, and WebAssembly are popular software packaging and distribution technologies. These technologies are leveraged in different abstraction layers to enable easier development and deployment of confidential computing applications. In the following, we give a brief introduction to these technologies. These approaches are summarized in Figure~\ref{fig:isolation_overview}, that illustrates the key elements and the boundaries of each isolation technology, alongside the mechanisms leveraged to interact outside the isolation perimeter.

\subsubsection{Virtual machines}

Virtual Machines (VMs) provide strong isolation by running full operating systems on virtualized hardware, making them a well-established solution for secure workload execution and multi-tenant environments. VMs offer strict isolation by leveraging hypervisors to separate workloads at the hardware level. While this comes with higher resource overhead and slower startup times compared to more recent technologies, VMs are the only technology that allows packaging and distribution of unmodified complex applications that are closely linked to a specific operating system.

\subsubsection{Linux-based containers}

Linux-based containers (containers for short) provide a lightweight and efficient approach to application deployment by encapsulating software and its dependencies in isolated user-space environments. Unlike traditional virtual machines, which require full operating system emulation, containers share the host Linux kernel while maintaining process-level separation through isolation mechanisms provided by the Linux kernel, such as namespaces and cgroups. This architecture enables rapid startup times, efficient resource utilization, and seamless portability across different Linux-based computing environments. However, applications unavailable on Linux or that require specific access to kernel features not available in a containerized environment will not be able to run in containers.

Containers have become the backbone of modern cloud-native applications, facilitating microservices architectures and scalable orchestration. Their flexibility and efficiency make them an essential component of modern software development and deployment strategies.

\subsubsection{WebAssembly}

WebAssembly (Wasm)~\cite{wasm_spec} is emerging as a lightweight and secure alternative to traditional Linux-based containers, offering a portable execution environment that delivers near-native performance. Unlike containers, which rely on OS-level isolation, WebAssembly runs within a small, highly sandboxed, virtual machine, enforcing strict memory safety and reducing the attack surface. This makes it particularly well-suited for untrusted code execution, edge computing, and multi-tenant environments.

Another advantage of Wasm is its portability, which allows applications to be compiled once and executed across different platforms without modification, unlike hardware-accelerated VMs or containers. This greatly helps the application developers, who do not have to distribute and test multiple binaries. Its small footprint and fast startup times make it an attractive option for computing on constrained devices, for example, in the far edge.

By default, Wasm applications are not able to interact with their outside environment because of sandboxing. The WebAssembly System Interface (WASI) is a modular set of system APIs designed to provide Wasm programs with secure access to resources such as files, clocks, pseudo-random number generators, and other resources.

WASI is capability-based, which means programs must be explicitly granted access to a specific resource. This enhances security by granting minimal privileges to Wasm applications by default. The WASI standards aim to be platform-independent, to allow Wasm applications to run consistently across operating systems and devices. WASI interfaces are designed with security in mind to ensure the integrity of the Wasm sandbox.

\section{Trusted execution environments}
\label{sec:tee}

Trusted Execution Environments are the main components of confidential computing, tasked with executing the enclave application and ensuring the protection of the data while in use. This section presents the main existing TEE technologies within the scope described in the introduction. They are classified by their design approach, and their inner workings are briefly described.

\subsection{Introduction to TEE design approaches}

The design of a TEE is an important factor in the usability and simplicity of use for application developers. In some cases, the design can even restrict which applications are able to run on the TEE (e.g., multiprocess, access to dedicated hardware). Two main approaches emerge from the ecosystem which are modeled on the common application units used in software engineering. These two units are a process (for a single, standalone software) and a VM (for more complex applications requiring software and dependencies to work tightly together). The TEE using these approaches are designated as App-based TEEs and VM-based TEEs, respectively.

Other, less common approaches also exist. Most of them are coming from research projects without broad use or industry support. The notable exception is Arm TrustZone~\cite{armtrustzone}, which is one of the first TEE technologies and is commonly used in embedded and mobile applications. However, these approaches can fill specific use cases or tackle new challenges that current TEE technologies have not yet addressed.


\subsection{Hardware security mechanisms}

TEEs are built upon a specific CPU architecture, itself based on a specific Instruction Set Architecture (ISA). There are various ISAs that are commonly used in the computing ecosystem relevant for this paper. Most of these ISAs are proprietary (such as x86 or ARM), and the CPUs leveraging those ISAs are often proprietary too. This makes security research on these ISAs complex since the researchers have no solution but to reverse engineer the CPUs to explore their security features. This explains the lack of information on the hardware mechanisms of CPUs based on those ISAs.

A notable exception to this ecosystem is the RISC-V ISA. This ISA is open and belongs to a non-profit that allows anybody to use the RISC-V specifications to build CPUs. This results in an ecosystem where the security solutions are often designed openly, or even directly included in the RISC-V specifications. As such, we are able to explore the RISC-V security mechanisms deeper than those of other solutions.

There are multiple hardware security mechanisms inside the RISC-V specifications that are often used to build TEE solutions~\cite{lu2021riscvhwsurvey}. The first one is the Physical Memory Protection (PMP). It allows to define a finite (up to 16, which is the maximal number of PMP registers in the specification) number of memory regions on which access permissions are defined. The permissions are enforced by the hardware implementing PMP.

Another mechanism (that is not yet standardized) aims at enabling multiple hypervisors running in parallel, an essential building block for VM-based TEEs or RISC-V. It is the Supervisor Domain Access Protection. Its goal is to add the support for physical address space isolation, thus allowing for OSes to run securely in parallel.

\subsection{Application-based TEEs}

Application-based (app-based) TEEs aim to secure a single process in an untrusted \gls{os}. They aim to protect designated chunks of memory inside a process. These TEEs generally allow for fine-grained control of the application memory, as well as the underlying TEE mechanisms. Most of the time, this control is made available through an SDK. The main benefit of this design is a small \gls{tcb}, as the developer explicitly chooses what data to store into the TEE. As the TEE is protecting the minimal amount of memory required, this approach is also generally the fastest one.

However, this comes at a heavy development cost, as existing applications must be adapted to work on the underlying TEE. Furthermore, the security-relevant variables must be correctly identified and secured using the SDK. A wrong use of the SDK could result in the leak of variables into the untrusted host, defeating the purpose of the TEE. Thus, the developers need to be able to accurately identify which memory parts of their code need to be secured. Optimization of the SDK use also involves a performance aspect, as the interactions between classic and protected memory need to be optimized in order to avoid an important performance overhead.

\subsubsection{Intel SGX}

Intel Software Guard Extensions (SGX)~\cite{intelsgx} is one of the first TEE technologies, introduced in 2015. The first version of SGX (SGXv1) was initially deployed in Intel's 6th-generation Core processors. In SGXv1, the CPU enforces access control for the enclave memory, known as the Enclave Page Cache (EPC). However, SGXv1 had several drawbacks. The enclave memory was limited (approximately 128MB EPC shared among all enclaves), it lacked support for dynamic memory allocation (pages could not be added after enclave creation), and multithreading support was constrained and difficult to manage securely.

To facilitate application development for SGX, Intel provides a Software Development Kit (SDK) that simplifies the complexity of enclave creation and management. The SDK includes a set of tools, libraries, and example code for building enclave-based applications, and is primarily targeted at C and C++ developers.

SGX's threat model assumes that attackers may control the OS, hypervisor, BIOS, and even certain hardware components, but cannot breach the CPU package. It also explicitly excludes side-channel attacks. Many academic studies have demonstrated that enclaves are vulnerable to various cache, page-fault, and speculative execution attacks.
Furthermore, SGXv1 suffered from various attacks weakening the solution~\cite{nilsson2020surveypublishedattacksintel}, among which were hardware-based attacks that are impossible to patch (even if mitigations are available).

Intel announced SGXv2 in 2019 to address several of the practical limitations of SGXv1 and improve the security of SGX. SGXv2 made enclave management more flexible and scalable for modern applications. Key enhancements include dynamic memory management (allowing enclave memory to be dynamically allocated and deallocated at runtime) and improved thread support.
Intel announced in 2022 that for the 11th and 12th generations of its processors, SGX would be deprecated for personal chips and that only the Xeon server processors would continue to support SGX.

\subsubsection{Keystone}

Keystone~\cite{lee2019keystone} is an open-source framework for building TEEs on top of the RISC-V instruction set architecture. It is designed to bring an App-based TEE to an open and extensible hardware platform. Keystone leverages the modularity of RISC-V to provide a flexible, formally verifiable TEE infrastructure that can be tailored to specific threat models and use cases.

Keystone introduces the notion of enclave-isolated execution contexts where code and data are protected from interference or inspection by any other software, including the operating system and hypervisor. It achieves this by using the RISC-V Physical Memory Protection (PMP). PMP allows the definition of memory access rules that prevent untrusted software from accessing enclave memory. These rules are managed by a minimal security monitor (SM). This small piece of trusted code running at the highest privilege level (M-mode in RISC-V) is responsible for enforcing isolation and managing enclave lifecycle events.

Keystone is significant for several reasons. First, unlike most x86 or ARM-based TEEs, which are proprietary and hardware-locked, Keystone is fully open-source and can run on open RISC-V hardware. This allows researchers and developers to design, inspect, modify, and formally verify the entire stack~\cite{cva6,fia_pmp_cva6}. It also allows modification of the solution for specific purposes. Second, Keystone targets standard RISC-V platforms, providing a vendor-neutral foundation for secure computing on a variety of general-purpose CPUs.
The simple requirements needed for Keystone that consist of a hardware Root of Trust (RoT), PMP support, and support for the basic U, S, and M modes allow for running enclaves on a large scale of RISC-V hardware, making confidential computing more broadly available than with proprietary alternatives that require specific CPU versions. A notable limitation of Keystone is that its maximum number of enclaves running in parallel is limited to $N-2$, with $N$ being the number of available PMP entries (as the SM consumes two entries for its own protection, and one entry is required for each enclave).

\subsubsection{Penglai}

Penglai~\cite{feng2021penglai} is an open-source RISC-V TEE that enhances enclave scalability and performance by addressing three key metrics: the size and granularity of secure memory, the number of enclaves, and the startup latency of the enclaves. The hardware security mechanisms standardized in the RISC-V specifications, such as PMP, are limited in these regards, notably regarding the flexibility and the number of enclaves. As such, Penglai solves this problem by proposing two hardware extensions: the Guarded Page Table (GPT) and the Mountable Merkle Tree (MMT). 

These custom hardware mechanisms extend the concepts of the standardized PMP while addressing its limitations. More precisely, GPT protects page table pages and enables memory isolation with page-level granularity, and MMT is a new abstraction to achieve on-demand and scalable memory encryption and integrity protection. Together, they remove the constraint of a fixed number of secure memory regions, and enable efficient management of a large number of enclaves. However, Penglai's reliance on GPT and MMT requires specific hardware support, making it less widely deployable than other RISC-V TEEs based only on PMP.

The system operates with a secure monitor in machine (M) mode and continues to rely on PMP for enforcing memory isolation. To reduce enclave initialization overhead, Penglai introduces shadow enclaves, which enable fast enclave instantiation by avoiding expensive memory clearing operations.

\subsection{Virtual Machine-based TEEs}

The other main approach, VM-based TEEs, aims to protect complete virtual machines instead of only one process. Thus, they allow for protecting complex applications or even multiple, independent applications at once. Their main drawbacks are a much larger \gls{tcb}, as the trusted code includes the entire \gls{os} of the VM, and a generally slower performance because of the virtualization~\cite{perflossvirt}. It is worth noting that, for most solutions, the TEE is built upon the virtualization acceleration solution of the hardware, that allows not to sacrifice too much performance. Their main benefit is that applications can benefit from the security provided by confidential computing without the need for modification. This approach allows for a quicker popularization of the use of confidential computing in various sectors.

\subsubsection{AMD SEV}

AMD Secure Encrypted Virtualization (SEV)~\cite{amdsev} was announced in 2016 and is the first TEE solution for virtual machines. SEV enables the memory of each VM to be transparently encrypted with a unique key managed by a dedicated security processor known as the AMD Secure Processor (ASP). This ensures that only the VM itself can decrypt its memory contents, preventing unauthorized access even by privileged system software running on the host machine.

The security of SEV has been improved throughout the years with extensions, mainly in answer to several attacks~\cite{werner2019severest,hetzelt2017sevanalysis}. In 2017, SEV-ES (Encrypted State) added encryption and integrity protection to the CPU register states during VM exits, preventing hypervisors from inspecting or manipulating VM execution contexts. In 2020, SEV-SNP (Secure Nested Paging) introduced hardware-based memory integrity protection, strengthening defenses against a broader range of attacks, including replay and remapping attacks, by validating guest memory accesses through a trusted memory encryption engine. Finally, AMD announced SEV-TIO (Trusted Input/Output) in 2023, allowing the establishment of trust and communication between trusted devices.

\subsubsection{AWS Nitro Enclaves}

AWS Nitro Enclaves (NE)~\cite{aws-nitro-enclaves} is a confidential computing feature offered by Amazon Web Services (AWS) that enables the creation of isolated, hardened virtual machines within EC2 instances (AWS' virtual machine solution). Built on top of the AWS Nitro System~\cite{security-design-of-aws-ne}, which is the underlying hardware and software platform for modern EC2 (VM) offering, Nitro Enclaves reserve a portion of the instance's CPU and memory to create a separate virtual machine with a reduced attack surface, communicating with the host instance only through a secure virtual socket interface. Applications are converted from an OCI container image to an enclave image using the \texttt{nitro-cli} command utility.

Unlike other TEEs, which are tied to specific CPU capabilities and enclave instructions, Nitro Enclaves rely on dedicated Nitro hypervisor technology to carve out secure memory regions within an EC2 instance. These enclaves have no persistent storage, no external networking, and no interactive access, minimizing the attack surface. Nitro Enclaves are available on Intel, AMD, or AWS Graviton (ARM-based CPUs developed by and for AWS) instances, making them theoretically cross-platform. However, they are deeply integrated into the AWS ecosystem and unavailable outside of it.

Although it is not clear which part the dedicated Nitro hardware takes in the protection of Nitro Enclaves, they may be considered a form of hardware-backed TEE, as AWS claims the isolation is enforced by the Nitro System at the hypervisor and memory-controller level. We choose to include Nitro Enclaves in this study as it is frequently mentioned in the confidential computing ecosystem and is generally considered as a TEE by the community.

\subsubsection{IBM Secure Execution}

IBM Secure Execution (SE)~\cite{borntrager2020ibmse} is a TEE specifically designed by IBM for its offer of mainframes on its in-house z/Architecture ISA. Introduced with the IBM z15 and LinuxONE III systems, Secure Execution provides strong confidentiality and integrity guarantees by creating a secure, encrypted enclave for each virtual machine, known as a Secure Virtual Server. 

Under IBM Secure Execution, a VM is first prepared offline. The tenant uses SE-specific tooling to encrypt the guest kernel, initramfs, and boot parameters, and to bind the resulting image to the target system’s SE keys. This ensures that only the platform’s trusted firmware (the ultravisor) can decrypt it. When the VM is started, the standard hypervisor delegates launch control to the ultravisor, which verifies the image metadata, decrypts the payload, and instantiates the guest inside a hardware-enforced protected address space. During execution, all guest memory is isolated from the host and hypervisor: pages remain encrypted when stored externally, access by host software is blocked, and integrity protections prevent tampering. As a result, the VM runs with confidentiality and integrity guarantees against both a potentially malicious host \gls{os} and administrative users, while preserving normal VM lifecycle operations.

SE only supports Linux, as its architecture is based on KVM. It is supported by Linux distributions certified for IBM Z platforms, such as Red Hat Enterprise Linux, SUSE Linux Enterprise Server, and Ubuntu.

\subsubsection{Arm CCA}

Arm Confidential Compute Architecture (CCA)~\cite{armcca} was introduced in 2021 along with the Armv9 architecture. CCA is a key feature of the Armv9-A architecture and makes confidential computing available to a wider audience for ARM machines. The first Arm TEE, TrustZone (presented in \ref{sec:trustzone}), was limited, and developing for the TrustZone TEE requires a strong collaboration with Arm and acceptance of related collaboration terms. Arm CCA represents a significant evolution in secure computing for the Arm ecosystem, offering a scalable and flexible path to deploy confidential workloads, compared to its limited TrustZone predecessor.

At the core of Arm CCA is the concept of Realms, a new form of secure execution environment for protecting virtual machines. Realms are instantiated and managed by a Realm Management Monitor (RMM), a privileged firmware component that operates beneath the hypervisor. Realms execute in a dedicated CPU mode and memory space that is cryptographically isolated using Arm's Memory Tagging Extension (MTE) and Secure EL2 mechanisms, along with a hardware-based component known as the Realm Management Extension (RME).

Arm CCA itself does not provide any implementation of an RMM. Islet~\cite{islet} is an open source implementation of the Realm Management Monitor (RMM) written in Rust. Islet is backed by the Arm company and other industrial actors and is considered as a reference implementation of the CCA RMM. Islet provides an open source SDK written in Rust. The Islet SDK enables developers to create applications for ARM CCA enclaves, offering built-in support for essential security primitives such as data sealing and secure channel establishment.  

\subsubsection{IBM PEF}

The Protected Execution Facility (PEF)~\cite{hunt2021pef} is IBM's TEE designed for confidential computing on the Power ISA, introduced with the POWER9 architecture. The Power ISA is developed by a foundation led by IBM, and is used in many IBM products. PEF enables the creation of secure virtual machines (SVMs).

SVMs are managed by a minimal trusted firmware component, called the ultravisor. The ultravisor controls the transitions into and out of secure state, enforces access control policies, and manages the cryptographic protection of memory belonging to secure VMs. Memory pages belonging to a secure VM are subject to hardware-enforced encryption and integrity checks before they leave the secure state or are paged out to normal memory. The ultravisor ensures that any page movement from secure to non-secure requires encryption (confidentiality) and cryptographic integrity (detection of tampering) before being made accessible to the hypervisor or system software. In turn, when those pages are later brought back into secure state, they must pass the integrity check and be decrypted before guest access.
\subsubsection{Intel TDX}

Intel Trust Domain Extensions (TDX)~\cite{inteltdx} is a hardware-based confidential computing technology introduced by Intel in 2023 on its 4th generation of Intel Xeon processors to provide enhanced security for virtualized environments. It enables the creation of Trust Domains (TDs), which are hardware-isolated virtual machines.

At the core of Intel TDX is the Secure Arbitration Mode (SEAM), a new CPU mode that hosts the TDX module responsible for managing TDs. This module operates in a reserved memory region defined by the SEAM Range Register (SEAMRR) and functions alongside the existing virtualization infrastructure. Memory confidentiality and integrity for each Trust Domain are enforced using hardware mechanisms. TDX leverages Intel Total Memory Encryption, Multi-Key (TME-MK), so that each TD’s memory is encrypted under its own key.

\subsubsection{RISC-V CoVE}

The RISC-V Confidential VM Extension (CoVE)~\cite{sahita2023coveconfidentialcomputingriscv}, previously named Application Processor Trusted Execution Environment (AP-TEE), is an open standard designed to enable confidential computing on RISC-V platforms. It first appeared in 2022 and is in the process of being standardized as an official RISC-V extension. It provides a hardware-assisted framework for running TEE Virtual Machines (TVMs) that are isolated from untrusted software components, including the host operating system and hypervisor. This isolation is achieved through a combination of RISC-V privilege modes, new Supervisor Binary Interface (SBI) extensions, and a minimal security monitor known as the Trusted Security Manager (TSM).

CoVE necessitates several hardware extensions, such as the H-mode for enabling hardware virtualization support, Supervisor Domain Access Protection for isolating physical memory between different supervisor domains (a more complete and flexible approach than PMP, as PMP is insufficient for creating a VM-based TEE), and Advanced Interrupt Architecture (AIA) for fine-grained interrupt control. Some of these extensions are still in the process of being officially standardized.

While CoVE lays the groundwork for secure VM-based execution on RISC-V systems, its specifications are not completely ready for general use and no hardware exists yet for some extensions needed by CoVE. However, it is a significant development in the TEE ecosystem, making it relevant enough to be included in this study. Furthermore, a research project by IBM, named Assured Confidential Execution (ACE)~\cite{ozga2025ace} implemented CoVE's specification by emulating hardware features that are not yet available. This project also works on the provability of a TEE Security Monitor~\cite{ace_formally_verified} (the trusted code responsible for managing enclaves and enforcing isolation).

\subsection{Other TEE approaches}

Some TEEs are using other approaches that do not fit into the application or virtual machine model. For the case of TrustZone, the approach is unique because, as the first TEE implementation, it made choices that were not reused for further iterations.
Finally, HyperEnclave proposes a flexible design based on both VM- and App-based TEEs, without requiring specialized hardware. 

\subsubsection{Arm TrustZone}
\label{sec:trustzone}

Arm TrustZone~\cite{armtrustzone}, introduced in 2004, represents Arm's first approach to hardware-based isolation. It is implemented in both the Arm Cortex-A and Cortex-M processor families, starting from ARMv6. It establishes two distinct execution environments, the secure world and the normal world. Each world has its own operating system, generally a rich operating system for the normal world, and a specialized secure operating system running in the secure world. The secure operating system governs the secure world and is responsible for managing access to protected resources. 

TrustZone supports only a single secure world, with a fixed secure operating system. This limitation makes it unsuitable for use cases involving multiple mutually distrusting workloads or dynamic provisioning of secure services. Additionally, TrustZone lacks support for generic virtualization, a feature only introduced later with Armv8.4-A through the addition of Secure EL2 (S-EL2), accessible solely from within the secure world. These limitations have motivated the development of more flexible secure computing architectures, such as Arm's Confidential Compute Architecture (CCA). 

Despite its constraints, TrustZone has seen widespread deployment in consumer devices. It has been used to enhance the security of Android smartphones, with platforms like Samsung Knox~\cite{samsungknox} leveraging TrustZone to protect sensitive operations and data. 

TrustZone on its own provides hardware isolation to create a TEE, but no software to leverage it. OP-TEE\footnote{https://www.trustedfirmware.org/projects/op-tee} is a widely used open-source runtime environment that is often used as the software layer on top of TrustZone. It provides an environment for creating trusted applications. OP-TEE adheres to the TEE architecture and the Global Platform API standard (GP API)\footnote{https://globalplatform.org}, which is structured around three main components: a client application, a specialized Linux driver, and the OP-TEE \gls{os}. In this setup, the operating system in the normal world is also known as the rich execution environment (REE). Host applications operate within this normal world and act as clients to trusted applications (TAs) in the secure world. These host applications use client APIs for communication.





\subsubsection{HyperEnclave}

HyperEnclave~\cite{jia2022hyperenclave} is a TEE technology that aims to overcome the limitations of existing vendor-specific TEEs like Intel SGX or AMD SEV. Instead of relying on specialized hardware features, HyperEnclave leverages standard virtualization extensions available on commodity processors to provide secure and isolated execution environments. It achieves this by running enclave processes inside lightweight virtual machines managed by a privileged security monitor called RustMonitor. This monitor, written in Rust, enforces strong isolation between the enclave and the potentially compromised operating system, ensuring that sensitive data and code remain protected. 

A key advantage of HyperEnclave is its cross-platform design. It can run on x86 systems from different vendors without hardware-specific dependencies. The system also maintains compatibility with Intel SGX applications, allowing existing enclave code to be reused with minimal changes. This makes HyperEnclave a mixed approach between VM-based and app-based TEEs, as it relies on VMs for the isolation but supports the app-based approach of SGX.






\subsection{TEEs takeaways}

The key findings of this section are that today’s TEE ecosystem remains highly fragmented, with vendors adopting divergent abstractions, programming models, and trust assumptions. An overview of TEE technologies presented in this section is available in Table~\ref{tab:base_hw}. Two dominant approaches have emerged: application-based TEEs, which isolate individual programs or functions within lightweight enclaves, and VM-based TEEs, which protect entire virtual machines using hardware-enforced boundaries. 

Application-based designs offer fine-grained isolation and smaller attack surfaces, but they often require substantial re-engineering effort and expose developers to platform-specific constraints. In contrast, VM-based solutions provide stronger compatibility with existing software stacks and simplify deployment, though at the cost of larger trusted computing bases and potentially higher overhead. While application-based TEEs offer better guarantees from a security point of view, VM-based approaches took over the cloud ecosystem thanks to their better usability. However, for other ecosystems such as the edge or embedded, the app-based TEEs remain essential because of their lightness. This dichotomy, along with other approaches that are also present in the ecosystem, underscores the lack of convergence in the field and highlights the need for unified abstractions and tooling to support secure and practical use of TEEs at scale.

\begin{table}[]
    \centering
    \caption{Overview of Trusted Execution Environments. The \emph{Year} column refers to the availability of technology and not its announcement date, except for CoVE where no hardware is currently available.
\label{tab:base_hw}}
    \begin{tabular}{cccccc}
        \bf Technology & \bf Acronym & \bf Year & \bf Hardware & \bf Approach & \bf Maturity \\
        \hline
        Arm TrustZone~\cite{armtrustzone} & TZ & 2004 & ARM & Other & Industry \\
        Intel Software Guard Extensions~\cite{intelsgx} & SGX & 2015 & x86 & App-based & Industry \\
        AMD Secure Encrypted Virtualization~\cite{amdsev} & SEV & 2016 & x86 & VM-based & Industry \\
        Keystone~\cite{lee2019keystone} & KS & 2019 & RISC-V & App-based & Research \\
        AWS Nitro Enclaves~\cite{aws-nitro-enclaves} & NE & 2020 & x86, ARM & VM-based & Industry \\
        IBM Secure Execution~\cite{borntrager2020ibmse} & SE & 2020 & z/Architecture & VM-based & Industry \\
        Arm Confidential Compute Architecture~\cite{armcca} & CCA & 2021 & ARM & VM-based & Industry \\
        IBM Protected Execution Facility~\cite{hunt2021pef} & PEF & 2021 & PowerPC & VM-based & Industry \\
        Intel Trust Domain Extensions~\cite{inteltdx} & TDX & 2021 & x86 & VM-based & Industry \\
        Penglai~\cite{feng2021penglai} & PG & 2021 & RISC-V & App-based & Industry \\
        HyperEnclave~\cite{jia2022hyperenclave} & HE & 2022 & x86 & Other & Research \\
        Confidential VM Extension~\cite{sahita2023coveconfidentialcomputingriscv} & CoVE & 2022* & RISC-V & VM-based & Research \\
    \end{tabular}
    \vspace{1em}
    \newline
    \small
\end{table}

\section{Abstraction layers}
\label{sec:al}

Abstraction layers are software solutions built on top of one or several TEEs to ease some aspects of developing and deploying TEE-protected applications. A more detailed context is given before presenting the abstraction layers ecosystem, categorized by the approach chosen for the software abstraction.

\subsection{Introduction and definition}

While TEEs provide strong hardware-enforced isolation for secure computation, their low-level nature and disparate inner workings present significant challenges for developers and system integrators. Programming directly with TEEs often requires expertise in cybersecurity, enclave development, and memory management, which can be complex and error-prone. Additionally, different TEE implementations have distinct APIs and security models, complicating portability and interoperability across platforms. 

To address these issues, additional software layers, that we will refer to as \textit{abstraction layers}, are needed to simplify the development and the deployment of confidential computing applications. These abstraction layers are software solutions that can provide additional features, including but not limited to multiplatform or multi-architecture deployment, support for additional programming languages, or deployment solutions. These features are aimed at improving the easiness of either the development or the deployment of TEE-protected application. Abstraction layers are built upon at least one TEE, but frequently support several of them.

This enables developers to focus on application logic rather than low-level security details, and makes confidential computing more broadly available for various applications. Furthermore, these abstraction layers can facilitate integration with existing software ecosystems, making confidential computing more accessible and reducing the risk of implementation errors that could compromise security.

\begin{figure}[!hptb]
    \centering
    \includegraphics[width=\linewidth]{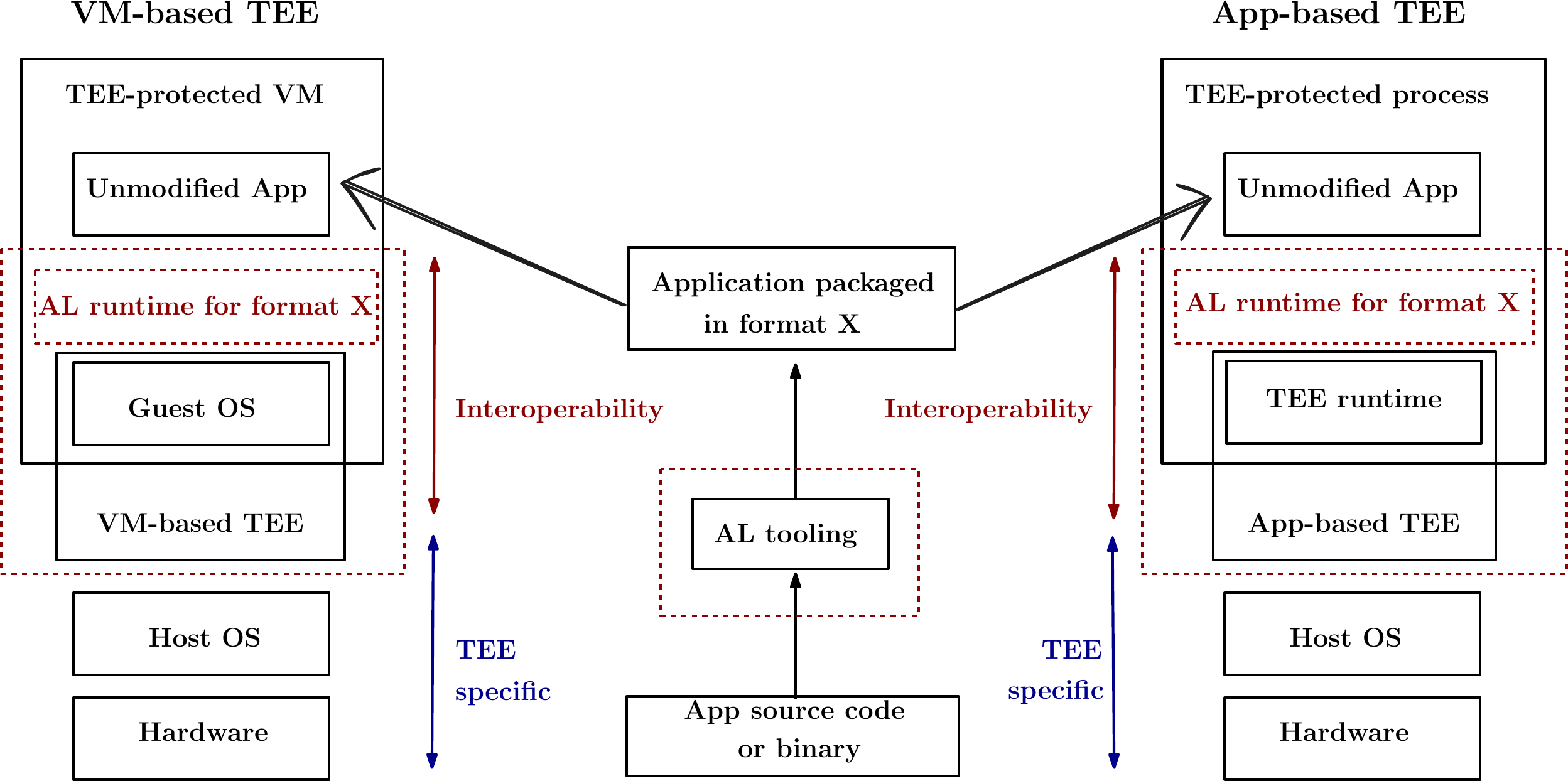}
    \caption{Representation of where the abstraction layer lays inside various TEE approaches. Dotted red represents the areas where abstraction layers are intervening. Blue arrows represent the classic TEE protection.}
    \label{fig:abstraction-layer-in-tee}
\end{figure}

The main goal of an abstraction layer is to enable the execution of an unmodified application, packaged into a designed format, inside a TEE. The interactions between an abstraction layer, a TEE (in the two main approaches), and an application are detailed in Figure~\ref{fig:abstraction-layer-in-tee}. An abstraction layer intervenes at two main steps of an application's lifecycle: first, when the application is being processed from a standard application to one supported by the abstraction layer. Second, on top of the supported TEEs, making the intermediary between the processed app and the TEE solution.

There are various approaches for abstracting a TEE. In the following, we detail the SDK-based, the container-based, and the WebAssembly-based approaches, as well as other, less common approaches.

\subsection{SDK-based abstraction layers}

SDK-based abstraction layers provide a software solution that simplifies the development of secure applications by exposing a high-level API and runtime environment, while relying on the underlying TEE hardware for isolation and protection. These solutions typically follow a client-enclave model, where an application executed on the untrusted \gls{os} communicates with a secure enclave (or trusted application) using well-defined interfaces. 

This approach is inspired by app-based TEEs that, for the majority, rely on an SDK to build an application that can run on said TEE. The main downside of the SDK-based approach is that applications still need to be developed for a TEE environment, without the possibility to port existing applications out of the box.

\subsubsection{Open Enclave SDK}

Open Enclave SDK~\cite{openenclave} is an open-source, hardware-agnostic SDK designed to unify and simplify the development of TEE applications across multiple TEE platforms. The SDK allows building applications using C or C++, and currently supports SGX and TrustZone. Developed originally by Microsoft, it provides a consistent, high-level programming model that abstracts away the hardware-specific details of each TEE, enabling developers to write portable and maintainable enclave applications.

The core functionalities of Open Enclave SDK mirror those of typical TEE SDKs: it supports enclave creation, secure communication between trusted and untrusted components, cryptographic services, and secure storage. However, it introduces several enhancements over using vendor-specific TEE SDKs directly. Most notably, it offers TEE portability, allowing a single codebase to target multiple hardware backends without modification. This is achieved through a unified API and a pluggable architecture that routes platform-specific behavior to appropriate backends at runtime or compile time. In addition, Open Enclave SDK improves the development and testing workflow with features such as a simulation mode, which allows enclaves to be executed and debugged without hardware support. 

\subsubsection{Asylo}

Asylo~\cite{asylo} is an open-source framework developed by Google to simplify the creation of applications running inside TEEs. It only supports Intel SGX, but the project was aiming to support multiple TEE solutions with its flexible design. Its design aims to abstract away low-level, hardware-specific complexities and offer a high-level, portable programming interface for secure applications.

Unlike TEEs that follow a split-world model, requiring developers to manually partition code into trusted and untrusted components with narrowly defined communication interfaces, Asylo supports a more unified approach. It enables the compilation of complete applications like Redis or SQLite into a TEE with minimal or no changes to the codebase, thanks to its support of the C and C++ standard libraries.

Internally, Asylo uses a sandboxed runtime and a pluggable backend system that allows the same application to run either in a real enclave or in a simulation mode for easier debugging. It also includes support for data sealing and enclave lifecycle management. Asylo is notable for its whole-application enclave model, which stood in contrast to the classic approach of modular and interface-driven SDKs found in the confidential computing ecosystem.

\subsubsection{Teaclave}

Teaclave~\cite{teaclave} is an open-source collection of developer-friendly SDKs, aimed at enabling direct development of custom TEE-protected applications. It was originally designed as a general-purpose secure computing framework for TEE-protected Function as a Service (FaaS) applications. This framework was relying on WebAssembly as its general-purpose execution target, but the Teaclave community valued direct SDK development because FaaS applications often have a very small codebase.

This collection of SDKs allow developers to use Rust, with an experimental SDK targeting Java. The Rust SDKs are currently supporting SGX and TrustZone, and the Java SDK supports SGX only, but the project aims to extend support to other TEEs.  

\subsection{Container-based}

Container-based solutions built on top of TEEs extend the familiar container abstraction to support TEE workloads. These solutions aim to combine the ease of deployment and isolation properties of containers with the strong hardware-backed security guarantees offered by TEEs. They allow users to run entire containerized applications inside a protected enclave. These solutions are based on widely used standards such as the OCI runtime spec~\cite{oci-runtime-spec} and image spec~\cite{oci-image-spec}. 

This approach features compatibility with standard container tooling, allowing developers to build and package applications using familiar tools. Combined with support for encrypted I/O and storage, this enables applications to securely process sensitive data without exposing it to the host. Some systems also integrate with Kubernetes, offering TEE-aware scheduling, and may provide enhanced logging, auditing, and policy controls. It provides an intuitive way to deploy and scale confidential workloads, abstracting away the low-level complexity of enclave programming while leveraging modern DevOps practices and hardware-enforced security.

\subsubsection{SCONE}

SCONE (Secure CONtainer Environment)~\cite{arnautov2016scone} is a secure container framework designed to run containerized applications inside SGX enclaves with minimal changes. First introduced in 2016, it is presented as a practical approach to securing cloud workloads using TEEs. 

SCONE targets containerized applications and enables them to run in an SGX enclave without requiring extensive refactoring. The core idea is to provide a transparent execution environment by integrating a custom libc inside the enclave, which becomes part of the \gls{tcb}. This libc intercepts and mediates all system calls, securely forwarding them to the untrusted host operating system using an asynchronous system call mechanism. This design allows applications to benefit from SGX protections without being aware of enclave-specific programming models. The SCONE libc also protects data at rest and in-transit by decrypting / encrypting data going through file descriptors, protecting files that are stored outside the enclave, and using TLS to secure communications.

To support efficient execution, SCONE also employs a kernel module on the host side to facilitate communication between the enclave and the OS, while maintaining confidentiality and integrity guarantees. This setup allows SCONE to handle I/O operations, signals, and thread management securely and efficiently, despite the limited functionality provided by the SGX hardware.

\subsubsection{Gramine}

Gramine~\cite{graphenesgx,tsai_libraryoses_2014,gramineproject} is a lightweight library \gls{os} designed to run unmodified Linux applications in a TEE. It provides a minimalist kernel-like environment that emulates Linux system services inside an enclave, enabling legacy applications to benefit from confidentiality and integrity protections without extensive rewriting. Gramine is primarily targeting Intel SGX, with recent work~\cite{graminetdx24} porting Gramine to TDX. Furthermore, this study mentions that the TEE-agnostic way the Gramine TDX porting is designed should allow for easy porting to other VM-based TEEs such as SEV.

Gramine operates by packaging the application together with a small runtime into a manifest file that specifies how the application interacts with the outside world. Inside the enclave, Gramine handles syscalls, file accesses, and I/O through a secure wrapper, ensuring that all interactions with the untrusted host \gls{os} are tightly controlled and, where necessary, encrypted or authenticated.

In addition to direct SGX execution, Gramine has a specialized tool called the Gramine Shielded Container (GSC) tool, which integrates with Docker to allow the conversion of regular OCI containers into SGX-enabled containers. This makes it easier to deploy confidential workloads using familiar container workflows without needing to manually manage the enclave configuration.

\subsubsection{libkrun}

Libkrun~\cite{libkrun} is a lightweight library designed to run processes inside a minimal virtual machine using KVM. Coupled with crun, a low-level container runtime developed by Red Hat, libkrun is able to run OCI-compatible workloads. Libkrun acts as a low-level runtime that enables strong isolation by encapsulating containers within virtual machines, while preserving compatibility with OCI container specifications through crun. Its goal is to offer a secure and efficient alternative to traditional container runtimes, especially in environments that demand stronger isolation than what classic Linux containers provide. 

A specialized variant, libkrun-sev, extends this model by integrating SEV support. With libkrun-sev, container workloads not only benefit from virtualization-based isolation but also gain confidential computing protection for SEV. Support is currently limited to SEV, and it is unclear whether future versions will include other TEEs. One notable improvement to libkrun-sev is its integration into Buildah, a popular container image-building tool also developed by Red Hat. A new command, \texttt{mkcw}, allows users to convert a standard OCI container into an SEV-protected confidential container, streamlining the process of deploying secure workloads without altering application code. 

\subsubsection{CoCo}

Confidential Containers (CoCo)~\cite{confidentialcontainerspaper} is an open-source project that brings confidential computing capabilities to the cloud-native ecosystem, enabling Kubernetes-based workloads to run with strong hardware-backed isolation. CoCo builds on earlier efforts from the Kata Containers project, which introduced lightweight virtual machines to improve container isolation by leveraging KVM-based virtualization. While Kata Containers provided an effective foundation, it became clear that its architecture was not suitable for creating a confidential computing framework supporting multiple TEEs. This realization led to the creation of CoCo as a broader and more modular solution for supporting diverse confidential hardware technologies. 

Internally, Confidential Containers orchestrate secure workloads by running each container inside a lightweight TEE-backed virtual machine. CoCo integrates with Kubernetes and container runtimes to transparently handle secure bootstrapping and secret provisioning, aiming to keep the developer and operator experience close to standard cloud-native tooling. A key goal of the project is to abstract the differences between TEE implementations, enabling developers to deploy confidential workloads without needing to adapt their code for each platform. CoCo currently supports SGX, SEV, TDX, and SE, with planned support for PEF.

\subsubsection{Anjuna Seaglass}

Anjuna Seaglass~\cite{seaglass} is a commercial confidential computing platform designed to simplify the deployment of secure applications using hardware-backed TEEs, with support for SGX, SEV, and NE. Seaglass supports deployment in containerized environments, including Kubernetes, using unmodified applications packaged in OCI containers. Its runtime stack handles the secure provisioning of secrets and secure I/O operations transparently. It also handles security policies, which govern how applications interact with sensitive data and ensure compliance with said policies throughout the lifecycle of a workload. These policies are enforced at runtime through integration with the platform's secure provisioning process. 

\subsection{WebAssembly-based}
In this section, we present \textit{Wasm Trusted Runtimes} that we consider in this study. A Wasm Trusted Runtime refers to a secure and isolated environment for executing Wasm modules in TEEs. This environment is often designed to ensure confidentiality, integrity, and attestation of computations.

A Wasm Trusted Runtime can be viewed as a \textit{two-way sandbox environment}, where the TEE provides hardware-level security, protecting the application from the host, while the Wasm runtime that is coupled with the TEE enforces software-level security, and protects the TEE and the host from the application. The following Wasm trusted runtimes and frameworks are developed to run untrusted or third-party Wasm code safely in a TEE.

Work on assessing Wasm runtimes has already been conducted in the past~\cite{zhang_research_2024,wang_how_2022}. While these surveys do not focus on trusted runtimes explicitly, they provide us with a base analysis of the runtimes that have been modified to make them trusted runtimes in the solutions presented below.

\subsubsection{Enarx}

Enarx~\cite{enarx} is an open-source framework that provides a Wasm trusted runtime. The current implementations (Enarx 0.7.1, January 2023)\footnote{https://github.com/enarx/enarx/releases} include support for Intel SGX and AMD SEV.

The Enarx project was initially conceived with the objective of developing a CPU architecture-independent runtime that facilitates application development without necessitating code modifications for compatibility across diverse hardware platforms. Just as other confidential computing platforms, the core of Enarx's design philosophy is to provide protection for data-in-use, referring to data residing within the processor during program execution.

Rather than asserting the trustworthiness of the host environment, Enarx adopts an architectural approach that minimizes the necessity to trust the surrounding system components. It does so by operating under the assumption that all entities external to the enclave (a \textit{keep} in Enarx's terminology) are inherently untrusted. In other words, only the enclave and the processor along with its firmware are trusted. Consequently, components such as the hypervisor, operating system kernel, and user-space services are denied access to both the application code and the data encapsulated within the enclave.

Enarx uses the following components in its architecture: (1) A Loader (in app-based TEEs) or a Virtual Memory Manager (in VM-based TEEs). (2) A Linux microkeonlyrnel (shim). (3) A Wasm runtime (Wasmtime). (4) A WASI Interface.

The goal of the Enarx components is to ensure that workloads are isolated and protected via an abstraction layer. Enarx uses WASI and the Wasm runtime, specifically Wasmtime, to provide APIs necessary for secure code execution and communication, that allows code portability across platforms without compromising security.

To securely execute a code in Enarx, an enclave is initiated on the host's machine. The following three steps are then executed: (1) \textbf{Authenticity Check.} Enarx verifies the authenticity of the application that the user intends to deploy. (2) \textbf{Packaging.} Then, Enarx uses cryptographic algorithms to encrypt the user's application and data. (3)
   \textbf{Provisioning.} Finally, application and data are executed in the enclave.

Enarx monitors code execution for abnormal behavior, such as unauthorized system calls or disallowed memory requests. If such behavior is observed, the enclave is terminated.

\subsubsection{TWINE}
TWINE (trusted Wasm in enclave)~\cite{menetrey2021twine} is an open-source, lightweight, and embedded Wasm virtual machine running in Intel SGX. It was developed by Ménétrey et al. in 2021. The authors expanded their work in 2023 and evaluated the performance of TWINE~\cite{menetrey2023comprehensive}.

TWINE provides hardware security by using the Intel SGX enclaves and software security by using Wasm to run applications, in a two-way sandbox environment. TWINE provides an environment to execute Wasm applications inside Intel SGX and consists of a Wasm runtime (that runs inside the enclave) and a WASI (that acts as the middleman between the runtime and the operating system).

In order to implement TWINE, several runtimes were evaluated as potential candidates. These runtimes were Wasmtime, Wasmer, Lucet\footnote{https://bytecodealliance.github.io/lucet/}, WAVM\footnote{https://github.com/wavm/wavm}, Wasm3\footnote{https://github.com/wasm3/wasm3} and WAMR. Most of these runtimes were excluded due to limited capabilities, such as large trusted computing base and therefore, more potential vulnerabilities, or lack of built-in TEE compatibility. Only WAMR met the requirements for a trusted WebAssembly runtime.

One of the reasons to use WAMR in TWINE is its "out-of-the-box" implementation capabilities for Intel SGX. The WAMR toolkit includes an ahead-of-time (AOT) compiler, allowing Wasm applications to be compiled into native code using LLVM before they are sent to TWINE's enclave. As a result, TWINE does not include a Wasm interpreter and is only capable of running AOT-compiled applications. The key advantage of this approach is that native code execution is faster than code interpretation. Additionally, the Wasm runtime has a smaller memory footprint compared to an interpreter, a vital consideration for SGX environments and cloud/edge computing scenarios. Embedding a just-in-time (JIT) compiler was not pursued in TWINE by its developers, because incorporating LLVM into an enclave would require adapting the code to meet SGX's restrictions. The implementation of TWINE has since been incorporated in the upstream WAMR project.


\subsubsection{WaTZ}
In 2022, a framework called WaTZ~\cite{menetrey2022watz} was proposed by Ménétrey et al. to securely execute Wasm codes inside Arm TrustZone. There is an overlap in scholars who developed WATZ and TWINE. WaTZ shares many design choices with TWINE, including the use of WASI, the reliance on WAMR, and the use of the Ahead-Of-Time (AOT) runtime model. However, WaTZ was developed independently of TWINE to adapt to TrustZone's unique design.

WaTZ has a significantly small footprint that reduces the attack surface, offering fewer opportunities for attackers to exploit compared to larger programs. WaTZ is specifically designed to provide a Wasm runtime for small, edge-scale Arm processors.


WaTZ uses the OP-TEE runtime. This utilization, however, had one drawback: OP-TEE's memory management API lacks the ability to adjust page protections to designate them as executable. To address this, the developers of WaTZ modified OP-TEE to enable such functionality for Trusted Applications.



\subsubsection{Other Wasm Trusted Execution Environments}

\paragraph{Se-Lambda}
Se-Lambda~\cite{qiang2018se} was developed in 2018 and uses Intel SGX to provide the two-way sandboxed environment. Se-Lambda provides serverless computing framework and remote authenticity check mechanisms. It was built on top of OpenLambda~\cite{hendrickson2016serverless}.

\paragraph{AccTEE}
AccTEE~\cite{goltzsche2019acctee} uses Intel SGX as its TEE. AccTEE runtime is open-source, and it offers a two-way sandboxing environment. AccTEE was developed in 2019, and as an “early” trusted runtime relied on the Node.js JavaScript runtime as a building component. Moreover, to execute JavaScript and Wasm, the developers of AccTEE used Chrome's JavaScript engine V8.




\paragraph{Veracruz}

Veracruz~\cite{veracruz} is a framework that provides a Wasm trusted runtime. Veracruz is focusing on VM-based TEE technologies for future development\footnote{https://github.com/veracruz-project/veracruz/issues/330}, such as Arm CCA and AWS Nitro Enclaves. Veracruz also uses WASI, as it provides a system interface for tasks such as accessing Veracruz's in-memory filesystem, generating random bytes, and similar operations. Veracruz presents a framework to develop Wasm applications in a TEE for privacy-preserving application scenarios~\cite{brossard2023private}. Wasmtime is chosen by the developers of Veracruz as the Veracruz Wasm engine.


\subsection{Other forms of abstraction layers}

\subsubsection{Constellation}

Constellation~\cite{constellation}, developed by Edgeless Systems, is a Kubernetes distribution designed to run on top of confidential computing technologies. It ensures that all data within a Kubernetes cluster remains encrypted at rest, in transit, and during processing, effectively shielding the entire cluster from the underlying infrastructure. 

The core of Constellation's security model involves running all Kubernetes nodes, including control plane and worker nodes, inside TEE-protected Virtual Machines using SEV or TDX. To maintain end-to-end encryption, Constellation transparently encrypts all network traffic between pods using WireGuard-based encryption and secures persistent storage, including state disks and cloud storage services. Cryptographic keys are managed within the CVMs, and key management is integrated into the cluster's lifecycle, ensuring that keys are securely generated, stored, and rotated without exposing them to the underlying infrastructure.

Constellation is a CNCF-certified Kubernetes distribution, ensuring compatibility with existing Kubernetes tooling and workflows. By integrating confidential computing into the Kubernetes ecosystem, Constellation enables a way to securely migrate sensitive workloads to the cloud while still leveraging the cloud-native ecosystem, especially around Kubernetes. 

\subsubsection{Occlum}

Occlum~\cite{shen2020occlum} is a memory-safe, multi-process library \gls{os} (an \gls{os} where functionalities are provided to the application as libraries, and compiled into an unikernel) designed to run applications within SGX enclaves. It enables unmodified or minimally modified Linux applications to execute securely inside SGX by providing support for compiling unmodified C or Rust code directly to a binary supported by Occlum. The build and deployment process in Occlum is inspired by common container tools, but Occlum does not support OCI images.

Traditional solutions built on top of SGX adopt designs that allocate a separate enclave for each process. On the contrary, Occlum adopts a single-enclave, multi-process architecture. This approach allows multiple processes to share the same enclave, significantly improving performance by reducing startup latency and inter-process communication overhead.

To maintain strong isolation between processes within the shared enclave, Occlum employs Software Fault Isolation (SFI)~\cite{sfi}. SFI is a software instrumentation technique that inserts runtime checks before every potentially dangerous operation, for sandboxing untrusted software modules within a single address space. This ensures that each process operates securely without compromising the integrity of others. More specifically, Occlum's SFI scheme is specifically adapted for SGX, so that multiple processes can live within a same enclave without interfering with each other.

Occlum is implemented in Rust, a memory-safe programming language, which inherently reduces the risk of common memory vulnerabilities inside the solution. The system supports various file systems, including read-only hashed file systems for integrity protection, writable encrypted file systems for confidentiality, and untrusted host file systems for convenient data exchange between Occlum and the host \gls{os}. 

\subsubsection{Oak}

Oak~\cite{google-oak} is an open-source platform designed to build distributed systems that are externally verifiable and rooted in hardware-based trust through TEEs. Oak is targeted at VM-based TEEs, and currently supports AMD SEV and Intel TDX. 

To minimize the size of the \gls{tcb}, Oak adopts a split architecture that separates applications into two components: an enclave application and a host application. The enclave component is executed within a TEE and handles security-critical operations, while the host component, which operates outside the enclave, serves as a frontend by exposing a gRPC endpoint. 

Oak offers two execution environments for enclave applications. The first is the Oak Restricted Kernel, a minimal microkernel designed to run a single application on a single CPU core. This option results in a minimal \gls{tcb} and provides strong isolation, but it requires careful partitioning of the application to isolate the trusted logic. The second option is the Oak Container, which embeds a full Linux kernel and a complete userspace environment, and supports OCI images.




\subsubsection{nitriding}

Nitriding~\cite{winter2023nitridingtoolkitbuilding} is an open-source toolkit designed to abstract the constrained development model imposed by Nitro Enclaves. While NE traditionally requires minimal network access and complex build processes, nitriding simplifies this workflow, allowing developers to run unmodified Linux applications inside enclaves with secure and seamless Internet connectivity. Additionally, nitriding emphasizes transparency and trust by making its entire codebase externally verifiable. 

To bridge the enclave's strict I/O constraints, nitriding provides a secure communication proxy that facilitates encrypted data exchange between the enclave and the external Internet via the untrusted host, without exposing the enclave to direct network access. This architecture decouples enclave development from the limitations of Nitro's native toolchain by supporting conventional application packaging, process management, and resource configuration.

\subsection{Abstraction layers takeaways}

 This section described an extensive list of abstraction layers. An overview is available in Table~\ref{tab:abstraction-layer}. Various approaches are used by these technologies, but three main ones can be identified. The SDK approaches propose a way of writing code that are compatible with multiple TEEs at once, or that are able to compile already existing applications without the need for modifications adapting them to TEEs. The other two main approaches, containers and WebAssembly, leverage well-known packaging and isolation technologies to execute software on various TEEs. 

 Abstraction layers can have very different features depending on what the abstraction layer goals are. Some of these features, along with discussion on how to implement them and their compatibility with the main approaches identified in this section, will be discussed in the next section.

\begin{table}
    \caption{Overview of abstraction layer technologies. The \emph{Year} column refers to the availability of technology and not its release date.
    \label{tab:abstraction-layer}}
    \centering
    \begin{tabular}{cccccc}
        \bf Technology & \bf Year & \bf Approach & \bf Supported TEEs & \bf Maturity & \bf Open source \\
        \hline
        SCONE~\cite{arnautov2016scone} & 2016 & Container & SGX & Industrial & \xmark \\
        Gramine~\cite{graphenesgx,tsai_libraryoses_2014,gramineproject} & 2017 & Container & SGX, TDX & Industrial & \cmark \\
        Oak~\cite{google-oak} & 2018 & Container, Other & SEV, TDX & Research & \cmark \\
        Open Enclave SDK~\cite{openenclave} & 2018 & SDK & SGX, TZ & Industrial & \cmark \\
        Se-Lambda~\cite{qiang2018se} & 2018 & WebAssembly & SGX & Research & \xmark \\
        Asylo~\cite{asylo} & 2018 & SDK & SGX & Research & \cmark \\
        AccTEE~\cite{goltzsche2019acctee} & 2019 & WebAssembly & SGX & Research & \cmark \\
        Occlum~\cite{shen2020occlum} & 2020 & Other & SGX & Industrial & \cmark \\
        Teaclave~\cite{teaclave} & 2020 & SDK & SGX, TZ & Industrial & \cmark \\
        Enarx~\cite{enarx} & 2021 & WebAssembly & SEV, SGX & Research & \cmark \\
        libkrun~\cite{libkrun} & 2021 & Container & SEV & Industrial & \cmark \\
        TWINE~\cite{menetrey2021twine} & 2021 & WebAssembly & SGX & Industrial & \cmark \\
        CoCo~\cite{confidentialcontainerspaper} & 2022 & Container & SEV, SGX, TDX, SE & Industrial & \cmark \\
        Constellation~\cite{constellation} & 2022 & Other & SEV, TDX & Industrial & \cmark \\
        Veracruz~\cite{veracruz} & 2022 & WebAssembly & NE, CCA & Research & \cmark \\
        WaTZ~\cite{menetrey2022watz}& 2022 & WebAssembly & TZ & Research & \cmark\\
        Anjuna Seaglass~\cite{seaglass} & 2023 & Container & NE, SEV, SGX & Industrial & \xmark \\
        Nitriding~\cite{winter2023nitridingtoolkitbuilding} & 2023 & Other & NE & Industrial & \cmark \\
    \end{tabular}
\end{table}

\section{Improvements brought by abstraction layers}
\label{sec:discussion}

In the previous sections, this study established a comprehensive and up-to-date database of abstraction layers, serving as a foundation for this section. Abstraction layers provide improvements to bare TEEs across several domains, which will be described and analyzed below, along with the tradeoffs involved in using an abstraction layer. The analysis of this section uses the underlying TEE as a baseline, focusing exclusively on features that the TEE cannot provide by itself.

\begin{figure}[!hptb]
    \centering
    \includegraphics[width=\linewidth]{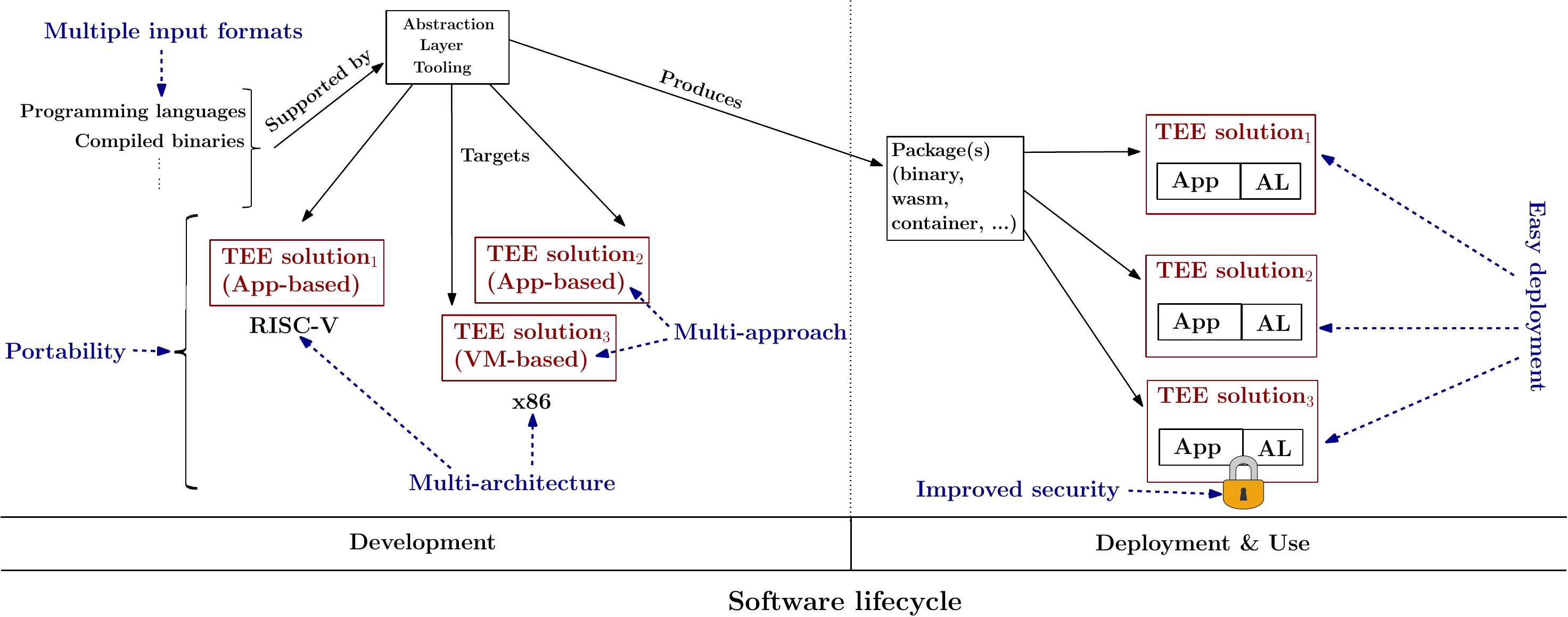}
    \caption{Representation of the lifecycle of a TEE-protected application leveraging an abstraction layer. Improvements proposed by abstraction layers are indicated using blue dashed arrows.}
    \label{fig:abstraction-layer-lifecycle}
\end{figure}

Abstraction layers intend to improve the development and deployment of confidential computing applications. These improvements can take place anywhere in the lifecycle of a confidential computing application, and may vary from an abstraction layer to another. Figure~\ref{fig:abstraction-layer-lifecycle} presents this lifecycle, that goes from the initial applications contents (binaries or source code, depending on the abstraction layers) to the final deployment of TEE-protected applications in various TEEs. It also highlights where the improvement brought by abstraction layers take place inside this lifecycle. Each improvement listed in Figure~\ref{fig:abstraction-layer-lifecycle} (multiple input formats, portability, easy deployment, and improved security) will be discussed in this section. Abstraction layers support for each of these improvements, along with other criteria, are indicated in Table~\ref{tab:al-improvements}.

This section begins by analyzing how different abstraction layer approaches can efficiently target the main TEE approaches. 
It then examines the improvements these layers introduce throughout the lifecycle of a TEE-protected application, focusing on its two main phases of its life: the development, and the deployment and use.
Finally, this section reviews how abstraction layers strive to preserve the security guarantees of TEEs.

\begin{table}
    \caption{List of improvement support for each abstraction layer.\label{tab:al-improvements}}
    \centering
    \begin{tabular}{c|cccccccc}
        \bf Technology & \bf Approach & \rot{\bf Multiple input formats} & \rot{\bf Portability}& \rot{\bf Multi-architecture} & \rot{\bf Multi-approach} & \rot{\bf Easy deployment} & \rot{\bf Improved security}& \rot{\bf Number of supported TEEs} \\
        \hline
        SCONE~\cite{arnautov2016scone} & Container & \ok & \nok & \nok & \nok & \nok & \ok & 1 \\
        Gramine~\cite{graphenesgx,tsai_libraryoses_2014,gramineproject} & Container & \ok & \ok & \nok & \ok & \nok & \ok & 2 \\
        Oak~\cite{google-oak} & Container, Other & \ok & \ok & \nok & \nok & \nok & \ok & 2 \\
        Open Enclave SDK~\cite{openenclave} & SDK & \nok & \ok & \ok & \ok & \nok & \nok & 2 \\
        Se-Lambda~\cite{qiang2018se} & WebAssembly & \ok & \nok & \nok & \nok & \nok & \ok & 1 \\
        Asylo~\cite{asylo} & SDK & \hok & \nok & \nok & \nok & \nok & \nok & 1 \\
        AccTEE~\cite{goltzsche2019acctee} & WebAssembly & \ok & \nok & \nok & \nok & \nok & \ok & 1 \\
        Occlum~\cite{shen2020occlum} & Other & \hok & \nok & \nok & \nok & \nok & \hok & 1 \\
        Teaclave~\cite{teaclave} & SDK & \hok & \ok & \nok & \nok & \nok & \hok & 2 \\
        Enarx~\cite{enarx} & WebAssembly & \ok & \ok & \nok & \ok & \nok & \ok & 2 \\
        libkrun~\cite{libkrun} & Container & \ok & \nok & \nok & \nok & \ok & \ok & 1 \\
        TWINE~\cite{menetrey2021twine} & WebAssembly & \ok & \nok & \nok & \nok & \nok & \ok & 1 \\
        CoCo~\cite{confidentialcontainerspaper} & Container & \ok & \ok & \ok & \ok & \ok & \ok & 4 \\
        Constellation~\cite{constellation} & Other & \ok & \ok & \nok & \nok & \ok & \nok & 2 \\
        Veracruz~\cite{veracruz} & WebAssembly & \ok & \ok & \ok & \nok & \nok & \ok & 2 \\
        WaTZ~\cite{menetrey2022watz} & WebAssembly & \ok & \nok & \nok & \nok & \nok & \ok & 1 \\
        Anjuna Seaglass~\cite{seaglass} & Container & \ok & \ok & \ok & \ok & \ok & \ok & 3 \\
        Nitriding~\cite{winter2023nitridingtoolkitbuilding} & Other & \ok & \nok & \nok & \nok & \ok & \hok & 1 \\
    \end{tabular}
\end{table}

\subsection{Approaches for an abstraction layer design}
\label{sec:approaches-al-design}

Implementing an abstraction layer requires working with the restrictions of TEEs and their various approaches. The abstraction layer approach will have an impact on its ability to target various approaches of TEEs. Furthermore, depending on the abstraction layer design, some may use strictly one enclave per application, where others will regroup several applications in one same enclave.

\subsubsection{Abstraction layers approaches targeting app-based TEEs}

The design of app-based TEEs are often aimed at running only one app per enclave. Therefore, it is difficult for an abstraction layer to leverage these TEEs in order to make them run multiple applications per enclave. This requires the development of a small operating system within the constraints of the TEE to be able to deploy multiple applications on top of it. The challenge of developing even a small secure operating system are well-known, and this \gls{os} will be added to the \gls{tcb} of the solution, making its security even more critical.

SDK-based abstraction layers, by design, only target one app per enclave, making their development on app-based TEEs relatively easy. Furthermore, most app-based TEEs are already providing their own SDK, allowing SDK-based abstraction layers to potentially reuse part or the totality of these SDKs.

Supporting app-based TEEs is way harder for container-based abstraction layers. Indeed, even if the goal of the abstraction layer is to run only one application per enclave, running a container requires a lot of effort. By design, the isolation of containers relies on the Linux kernel, the container needs a filesystem, the ability to run ELF files, etc. Even with a preprocessing step that adapts the container images before they are being executed, there is still a sizable amount of software that needs to be run into the enclave to support the features of containers. 

WebAssembly-based abstraction layers require a WebAssembly runtime that is running inside the enclave. Since WebAssembly is very lightweight, it is used to target small embedded devices. This resulted in the development of small, portable WebAssembly runtimes suitable for being integrated into an app-based TEE. The challenges described above about the execution of multiple binaries in parallel still applies to WebAssembly. However, the simplicity of its binary format and ISA, compared to more complex formats like ELF and x86, makes the development of such an operating system slightly easier.

\subsubsection{Abstraction layers approaches targeting VM-based TEEs}

While software aimed at VM-based TEEs is often portable due to standardized VM image formats, slight differences in functionalities (support for memory encryption, protection against memory corruption, etc.) exist~\cite{vm_cc_sok}. Such differences need to be taken into account while designing an abstraction layer that need to be portable across these TEEs.

VM-based TEEs are mainly aimed at running a whole virtual machine. While this approach is well-suited for public cloud ecosystems, where a virtual machine commonly serves as a unit of computing, it may not be optimal for more general use cases. Indeed, including a whole \gls{os} in the \gls{tcb} is not desirable from a security point of view. Abstraction layers may thus aim at reducing the \gls{os} size, or even removing the \gls{os} altogether and replacing it with a standalone application.

However, if the goal is to deploy a large number of applications protected by TEEs, it may be simpler from a system administrator point of view to include the deployment and orchestration solution inside the TEE directly, especially if said solution has sensitive data to protect on its own. Taking Kubernetes as an example, protecting its secrets and other sensitive content is not trivial and adds another layer of security on top of the TEE management, whereas integrating Kubernetes directly into the TEE removes this constraint. This approach is adopted by Constellation~\cite{constellation}.

From a cloud developer point of view, one of the main benefit of an abstraction layer is allowing them to easily, or even transparently, exploit the advantages of a TEE. VM-based TEEs can protect several applications in a single enclave. Furthermore, they can allow integrating a cloud solution (e.g., Kubernetes) within a TEE, enabling developers to transparently use features from the cloud solution while being protected inside a TEE.

Another argument in favor of embedding the orchestration solution inside the TEE is the number of enclaves. When deploying numerous applications with each its own enclave, the performance may be affected. The number of applications may even be capped, as most TEEs have a limit on the number of enclaves that can be run in parallel. By running the orchestration solution and all applications inside the same enclave, it reduces the performance overhead and removes the potential limitation on the number of applications. On the other hand, running all applications in the same enclave prevents from separating the applications from each other, which may be useful in the case of multi-tenant or multi-criticality environments.

This difference in approach stems from the same difference in philosophy that led to the app-based and VM-based TEE approaches distinction; either the emphasis is placed on fine-graining the security to each application, which may incur more costs in terms of design and development of the security solution, or the emphasis is made on the simplicity of use, gathering all applications and a large \gls{tcb} inside one enclave. Both have advantages and disadvantages. In this context, one also has to take into account not only the developers experience but also the system administrators experience.

\subsection{Improvements on development}

Abstraction layers can simplify the development process by providing higher-level programming models and APIs that conceal the complexities of low-level TEE interactions. This allows developers to focus more on application logic rather than on TEE specifics, to have faster development cycles and to reduce time-to-market for new applications. They allow a broader range of developers to engage with TEE technology, including those who may not have deep hardware or cybersecurity knowledge.

\subsubsection{Multiple input formats}

An abstraction layer may propose several possibilities for tooling to develop an application. These possibilities may be proposed alongside or on top of existing solutions coming from the TEEs underneath. Having more options for creating an application using an abstraction layer allows to leverage the skills of a bigger pool of software developers. This also lowers the barrier to accessing the benefits of confidential computing.

Furthermore, as observed in Section~\ref{sec:tee}, most TEE solutions that require an adaptation of the software for their solution provide an SDK that only supports the C programming language. C is a memory-unsafe language that is more and more criticized among the cybersecurity community, especially for the development of new applications. Abstraction layers bring the possibility of using newer, memory-safe languages as an alternative to C/C++, such as Rust or Go. For example, both these languages are able to compile static binaries used in containers. For WebAssembly, developers can use any programming language that compiles to Wasm via LLVM~\cite{lattner2004llvm} or other compatible toolchains, thus relaxing language constraints for secure execution.

Another possibility brought by abstraction layers is to consume already packaged applications (as compiled binaries, container images, or other formats) and adapt them for confidential computing use. SCONE is an example of such practice, where an off-the-shelf container image can be adapted to run on SGX. This allows to package not only in-house software but also well-known software packaged as container images (such as web servers or databases) to use along with said software. This improves the ease of development along with development times by allowing the leverage of tried and tested software.

Some abstraction layers are even able to run a binary not designed for running inside a TEE without any modification. Indeed, SCONE modifies the libc embedded inside the container to adapt it to SGX. This modification may introduce subtle changes in the behavior of the application that may be difficult to troubleshoot. Furthermore, reproducing such bugs requires providing the developer with access to corresponding hardware, which may not be easily available. Abstraction layers relying on WebAssembly and WASI such as TWINE~\cite{menetrey2021twine} are able to execute Wasm binaries targeting WASI without any modification. This brings a great testability, allowing the developer to test on its development machine and eliminating the need for additional, costly hardware. 

\subsubsection{Portability, multi-approach and multi-architecture}

An essential feature brought by abstraction layers is portability, i.e., the possibility to support multiple TEE technologies without having to rewrite TEE-specific code. This is relevant to avoid vendor lock-in, to improve the availability of an application, or to distribute it across different environments. Portability can be more or less flexible depending on how the abstraction layer is implemented. The assessment of an abstraction layer's portability can be split into two sub-criteria. The first one, multi-approach, is the ability to support multiple TEE design approaches as presented in Section~\ref{sec:tee}. The second one, multi-architecture, is the possibility of supporting multiple hardware ISAs. Not supporting one of these criteria limits the abstraction layer possibility of extending support to a large range of TEEs.

A multi-approach is relevant for developers aiming at targeting various computing environments with the same application. While VM-based TEE applications may be easily ported from one technology to another, doing so from a VM-based TEE to an app-based TEE or between two app-based TEEs is complex and often requires a rewrite of the application. Abstraction layers remove these constraints and allow using one codebase for all approaches.

Multi-architecture is an important feature for distributing a confidential computing application across various hardware targets. This is especially a challenge for edge applications, which are often required to run on more varied hardware platforms than the ones commonly found in clouds. The multi-architecture constraint breaks the portability of VM-based applications, as most VM-based platforms support only VMs that are using the same ISA as the host. If they can run VMs for other ISAs than the one of the host, it often comes at a huge performance cost because of emulation.

The SDK, container, and WebAssembly-based abstraction layers are all able to support each of the two main TEE approaches, although at different levels of efficiency, as already discussed in Section~\ref{sec:approaches-al-design}.
However, designing an abstraction layer that can efficiently support both approaches in a unified manner is more complex. As app-based TEEs are usually more constrained than VM-based TEEs, abstraction layers that aim to support both approaches often adopt designs that are constrained by app-based TEEs, before moving the design to VM-based TEEs. 

For a multi-approach abstraction layer, one of the most important part of its design is the definition of interfaces through which the application interacts with its environment. These interfaces must be as portable as possible in order to integrate well with all approaches. SDK-based abstraction layers are offering a lot of flexibility in this regard, as the interfaces provided by the SDK can be tailored by the abstraction layers creators. Containers are less flexible as they rely on various Linux mechanisms to interact with their environment, mechanisms that can be difficult to port to an abstraction layer. Finally, for WebAssembly, the use of WASI as interface allows to decouple application logic from specific TEE implementations. As long as a TEE (such as Intel SGX) can support or interpret Wasm modules conforming to WASI, applications can be executed securely across different enclave technologies, promoting broader TEE compatibility than containers. 

For the multi-architecture support however, containers and SDK fall short compared to the WebAssembly approach. Indeed, if containers can support multiple ISAs, one usually needs to create one container per ISA, multiplying the number of artifacts the developer has to maintain and test. SDK-based abstraction layers can in theory support a large number of ISAs, but in practice this large support is constrained by the design of the abstraction layer and the tooling used to transform the source code into something usable by the TEE. Indeed, providing robust support for a large set of ISAs, even for a single language, is not a simple feat, and abstraction layers not relying on a well-known compiler framework (such as LLVM), or regularly integrating the improvements of such a framework into the abstraction layer tooling, will fall behind WebAssembly on ISA support. Furthermore, SDK-based abstraction layers will still produce different binaries for different ISAs. On the other hand, WebAssembly-based abstraction layers produce one WebAssembly binary for all supported approaches and ISAs.

\subsection{Improvements on deployment and use}

Abstraction layers can ease the difficulty of deploying TEE-protected applications across multiple, potentially complex, environments. They can also help with improving the security guarantees offered by the TEEs.

\subsubsection{Easy deployment}

One of the key features of abstraction layers is to ease the amount of work to deploy TEE-protected applications in production. Depending on the use case, production deployments must take into account various criteria such as availability, monitoring, available hardware, and integration with existing solutions. This last point is essential since the integration of TEE-protected applications into an already existing infrastructure should be as smooth as possible. In this regard, SDK-based abstraction layers often rely on the deployment technologies provided by the underlying TEEs, and thus are difficult to integrate with other deployment technologies. On the other hand, both container and WebAssembly technologies are well-integrated with cloud-native deployment technologies. Abstraction layers based on them consequently allow for an easier management of distributed applications, bringing out-of-the-box lifecycle management, monitoring, and other features provided by orchestration platforms such as Kubernetes. 

WebAssembly goes even further thanks to its portability and lightness, that gives WebAssembly-based abstraction layers the possibility to be deployed on a large list of constrained devices. This allows system administrators to leverage TEEs present on edge and far-edge devices, which are becoming more and more relevant as the need to process sensitive data near their production points grows. Another point where abstraction layers ease the deployment of TEE solutions is by improving the interoperability between different TEE implementations. This enhances flexibility for system administrators to switch hardware or deploy across different environments without re-architecting their solutions.

However, while abstraction layers can provide numerous advantages when deploying an application in a TEE, they can also introduce drawbacks, such as performance overheads, complexity, layering issues, and dependency on the abstraction layer. Indeed, the abstraction will most of the time lead to additional latency or resource consumption compared to bare-metal execution. Furthermore, the additional layers that provide the abstraction layer capabilities can complicate the system architecture and potentially mask issues. Finally, an application's reliance on a specific abstraction layer's features may limit future flexibility and adaptability if this abstraction layer does not follow already existing and well-used specifications.

\subsubsection{Improved security}

While the abstraction layers must take care of preserving the security guarantees offered by the underlying TEEs, they can also augment the security of the application in several ways. First, the applications themselves can be verified for correctness (e.g., absence of undefined behavior) and protected (e.g., addition of security mechanisms) in order to avoid the compromission of the abstraction layer or the environment outside the TEE. For SDK-based abstraction layers, verifications can be made at compile time, but the code is often not isolated during execution.

In container- and WebAssembly-based approaches however, the underlying technologies are providing by design an isolation between the application and the abstraction layer. This isolation is especially welcome when the abstraction layer implies that several applications run in parallel within a single enclave. WebAssembly goes even further as some protections such as the Control-Flow Integrity (CFI)~\cite{cfi} can protect the application from malicious actors interacting with it.

By isolating the application from the abstraction layer and the underlying TEE, these approaches allow considering a stronger attacker model than the one commonly used for confidential computing and described in Section~\ref{sec:attacker-model}. Indeed, attackers coming from within the TEE are not considered in the classic confidential computing attacker model, while it can happen in practice (e.g., a TEE-protected vulnerable web service that gets compromised). Sandboxing technologies like WebAssembly or containers enable a new protection of the TEE host from the guest application, using a two-way sandbox. This is particularly interesting for abstraction layers that regroup several applications in one same enclave, as a compromised application will not necessarily result in a compromise of the totality of the enclave contents.

Abstraction layers on top of TEEs can improve security by constraining how applications interact with the TEE mechanisms. By construction, they provide a dedicated interface for TEEs, which helps to simplify and systematize the correct use of their low-level mechanisms. Since these layers are explicitly designed with TEE constraints and threat models in mind, they can enforce consistent and secure usage patterns and reduce the likelihood of developer errors. In addition, they can ensure that all TEE events are consistently handled according to the expected security model. In contrast, applications built directly on top of the TEE may overlook or mishandle some of these events, thereby introducing subtle but critical security weaknesses.

\section{The future of abstraction layers}
\label{sec:future}

Surveying the ecosystem of abstraction layers shows a large ecosystem with multiple solutions aiming at solving different problems. However, not all challenges are addressed by abstraction layers yet, and new challenges may arise in the near future.

\subsection{Comparing current approaches}

The three main approaches identified to build an abstraction layer are the SDK-based, the container-based, and the WebAssembly-based approaches. They all have their advantages and disadvantages. In addition, each abstraction layer is building on top of these approaches but does not necessarily implement all the features or follow the specifications of these base approaches, potentially breaking some advantages brought by them. However, in general, the approaches pros and cons that have an impact on the abstraction layer features can be identified and are summarized in Table~\ref{tab:al-approaches}.


\begin{table}
    \centering
    \small
    \caption{Comparison of the abstraction layers main approaches}
    \begin{tabular}{c|c|c|c}
         & \bf SDK-based & \bf Container-based & \bf WebAssembly-based \\
        \hline
        \bf Multiple input formats & only source code & support multiple, complex apps & high, lack expressiveness \\
        \bf Portability & no features & portable but with constraints & very portable and light \\
        \bf Easy deployment & no features & strong cloud ecosystem support & partial cloud ecosystem support \\
        \bf Improved security & some security mechanisms & strong sandboxing & very strong sandboxing \\
    \end{tabular}
    \label{tab:al-approaches}
\end{table}

WebAssembly stands out as the approach providing the best portability and security improvements. Containers are following close behind, but the fact that they are heavier than WebAssembly applications, less portable especially across ISAs, and requires a Linux kernel (or another solution implementing its isolation features) make them less ideal than WebAssembly, especially on edge devices. However, containers approaches remain the best choice for packaging complex sets of applications, since WebAssembly still lacks expressiveness for advanced features. Containers also benefit from a strong cloud ecosystem support, making them the best choice for targeting TEEs on public clouds. Finally, SDK-based approaches are lagging behind, mostly because they use an outdated design and mindset, that was relevant when confidential computing was a niche technology with a small number of applications designed explicitly for specific needs. The confidential computing ecosystem and its usages have grown in such a way that makes SDK-based abstraction layers mainly irrelevant for applications that are not explicitly targeting a TEE.

\subsection{Security challenges}

Even if abstraction layers added on top of TEEs can enhance security guarantees of TEEs, the effectiveness of the abstraction layers in preserving security can vary widely depending on the specific TEE technology and implementation and various other factors, including the design and the implementation of the abstraction layer itself, how it interacts with the underlying TEE, and potential vulnerabilities introduced by this additional layer. If the abstraction layer is correctly implemented, it can significantly contribute to the security of the application, offering both reinforcement of the security guarantees offered by TEEs, and additional security features for the application.

\subsubsection{Preservation of security guarantees}

One of the primary security guarantees of TEEs is the isolation of secure code and data from the host operating system and other applications. When an abstraction layer is introduced, it must maintain this isolation. If the abstraction layer inadvertently exposes sensitive data or allows improper access to the secure environment, this could compromise security. Proper access controls must be integrated within the abstraction layer to ensure that only authorized applications can communicate with or invoke secure operations within the TEE. If these controls are weak or improperly implemented, they can frame a potential breach of security guarantees. 

Abstraction layers can potentially increase the attack surface by introducing additional interfaces and pathways for interaction. Each additional layer can provide new opportunities for exploitation or new vulnerabilities if not designed and implemented carefully. As such, the robustness of the abstraction layer's design is critical for maintaining security. If the abstraction layer facilitates the transfer of sensitive data to and from the TEE, it must employ strong cryptographic measures to protect this data-in-transit. Weak data handling practices can expose sensitive information. For example, if the layer fails to verify the origin of requests or inputs intended for the TEE, it could lead to unauthorized access or manipulation of secure operations. A well-designed abstraction layer that incorporates layered security principles, such as segmentation, strong authentication, and regular security assessments, can help to preserve the security guarantees of the TEE. The security of the abstraction layer can also be supported by standardized practices and thorough auditing. Layers that are widely used, peer-reviewed, and subjected to rigorous security assessments are more likely to maintain the guarantees of the underlying TEE.

By design, an abstraction layer will increase the \gls{tcb}. The abstraction layer code that is being executed in the enclave needs to be trusted in addition to the TEE code that is executed in the enclave. While there is no guarantee that the abstraction layer and TEE code will be without any vulnerabilities, measures can be taken to minimize the risk, such as security audits. Another stronger possibility is to formally verify the code, as this has already been researched by the ACE study on CoVE~\cite{ace_formally_verified}. This verification work can be extended to any software of firmware that takes part of a TEE \gls{tcb}.

\subsubsection{Leveraging TEE features}

Different TEE implementations can support different sets of features (security or otherwise). These features can even be outside the scope of confidential computing, bringing additional protections against other attacks (e.g., supporting memory encryption to protect against hardware attacks). For an abstraction layer, supporting multiple TEE solutions may lessen the security offered by the TEEs. Indeed, when covering several solutions with different features it is difficult for the abstraction layer to cover them all while keeping one unified way of packaging the application. The abstraction layer can aim at covering all the features proposed by each supported TEE, which will result in an additional burden for the developer and complexity in the design of the abstraction layer. It may also require the abstraction layer to provide a way to abstract the support of features, meaning that the abstraction layer will have to provide a fallback in case a TEE does not support the feature used by the application. Another approach is to cover only the intersection of the features, but then the abstraction layer will fail to accurately exploit all the protection offered by each supported TEE. 

Furthermore, providing ways to abstract the TEE features may come with a security risk. Indeed, the application may want to know if a specific security feature is available or not and act in consequence. This goes against the idea of an abstraction layer providing all features transparently and without alerts, yet failing to provide the same level of security or efficiency. Indeed, for features such as memory encryption, the hardware-backed implementation provided by a TEE will often be more secure than the software-only implementation proposed by an abstraction layer as a failover. This security risk is even more present when the abstraction layers are used in conjunction with automatic orchestration or deployment solutions, because then the TEE-protected application may be redeployed on another TEE at any moment, without warning and without requiring an explicit approval from the system administrator.

In order to protect against the security risks posed by such discrepancies, abstraction layers should provide a way to list the characteristics of the TEE the application is currently running on, along with a way to be warned about a TEE change if the abstraction layer supports such a feature. Furthermore, for system administrators it is important to list the ways the TEE-protected application may be redeployed, on another TEE or not. If using automatic deployment or orchestration systems, it is important to check the parameters of the mechanisms that may trigger a redeployment, ensuring that such an action can only happen on the same TEE, or cannot happen without an approval or a warning to the system administrator. These requirements could be fulfilled by an application manifest, where the application can express which features are required and which ones are desirable, so the abstraction layer and the potential orchestrating mechanisms are able to avoid executing the application on a TEE where the required conditions are not met. The contents of this manifest could come from either a manual human source (the developers or the system administrators) or automatically generated from the application.

\subsubsection{Ensuring the continuity of data protection}

Confidential computing aims to protect data in-use, but doesn't guarantee the protection of this data when it is stored on disk or transported through a network. It is still up to the developer to ensure that the modification of the data state and its displacement are done securely. This is especially problematic for applications that are not designed for running inside a TEE, and therefore are potentially not handling file or network encryption. By allowing to execute generic applications in TEEs, abstraction layers become responsible for the continuity of the data protection, when it is being stored at-rest or transferred over the network.

An abstraction layer can help during the data transformation by proposing an integrated way to either store or transport data securely that has been produced by the application running inside an enclave. Furthermore, it can propose a transparent data encryption mechanism so that the developers don't have to think about data security at all. A transparent data encryption mechanism also makes it possible to leverage the protection of a TEE in an application that was not conceived for a TEE in the first place, by porting it as is into the TEE using the abstraction layer, and let the transparent encryption seamlessly protect the data. Ideally, to allow generic applications to benefit from the protection of TEEs while ensuring the complete protection of the manipulated data, an abstraction layer should be able to handle the continuity of both data at-rest and data in-transit transparently.

\subsection{Possible improvements for a unified confidential computing support}

The most notable contributions of abstraction layers to confidential computing as of today are to the execution environments. Therefore, this study is purposely scoped to the execution environment only and ignores other aspects of confidential computing. However, in the future, abstraction layers may be able to abstract not only the TEE solution but other aspects of confidential computing.  

\subsubsection{Unified, abstracted attestation}

One important aspect of confidential computing is attestation. Attestation allows providing a proof to the TEE beneficiary that its application is indeed being executed in a TEE. While the principle of attestation is the same across various TEE implementations, they differ enough in practice to make them incompatible between each other. Indeed, use of different cryptographic schemes, reliance on an external party or not, and discontinuous hardware support are various factors leading to a fragmented attestation ecosystem.

Recent efforts aim to create and standardize a unified attestation protocol for TEEs. The standardization of the Remote ATtestation procedureS (RATS) architecture~\cite{ietf_rats_RFC9334} or Weinhold et al.~\cite{weinhold_separate_2025} work on integrating remote attestation into TLS and comparing it with other existing protocols are examples of the interest in developing an attestation protocol that answers the needs of various TEE designs and user needs. However, as of today, most TEEs implement their own attestation solution with little to no support for these new unified protocols. This creates a gap that could be filled by attestation abstraction layers that could provide a unified attestation solution by uniformizing the existing solutions and merging them into a global, unified protocol. 

\subsubsection{Improved underlying technologies for abstraction layers}

In the technologies used in the main approaches, improvements are also possible. Improving the base technologies may result in an improvement on abstraction layers based on said technology, and open the way to new abstraction layers implementations leveraging these new features. Notably, the WebAssembly ecosystem is in a huge transformation, with a major release of the WebAssembly spec that occurred in September 2025, and constant improvements being made in the WASI specifications, resulting in WebAssembly binaries gaining more and more capabilities. This specification and standardization effort is followed by important work on the compiler and developing toolchains, making compiling complex code to WebAssembly increasingly easier. In this context, we may see appear abstraction layers that are supporting WebAssembly and WASI specifications and allow to transparently use a WebAssembly binary inside a TEE, while some WebAssembly-based abstraction layers require recompilation of the code with their custom compiler as they do not entirely support these standards.

The evolution of the container ecosystem is slower but still happening. The main improvements are taking place in the cloud ecosystem, especially in the domains of deploying, orchestrating, and monitoring containers. In this context, there is a possibility of developing abstraction layers that seamlessly integrate TEE-backed applications alongside regular ones, allowing application developers and cloud engineers to exploit the advantages of confidential computing while retaining their already existing cloud environments. New ways of expressing the need for applications targeting confidential computing, such as a manifest expressing the needs of said applications, may be integrated into the existing container standards. This standardization effort would allow the container and cloud ecosystem to fully embrace confidential computing for regular workloads, reaching a new height in data protection and privacy standards.

Finally, SDK-based abstraction layers are constantly being improved by the programming languages ecosystems they rely on. Languages and compilation toolchains improvements result in more secure and more performant binaries being produced. New languages with features relevant for the confidential computing ecosystem can also emerge and be adopted by existing or new SDK-based abstraction layers.

\subsubsection{Alternative methods of protecting data-in-use}

Confidential computing is competing with other technologies such as Fully Homomorphic Encryption (FHE) for protecting data-in-use. While FHE is vastly different from confidential computing, in some cases applications may benefit from both technologies at once, or a solution may require to support both confidential computing and FHE. In this context, abstraction layers that support both solutions may appear, leveraging one or the other when appropriate or porting a subset of what is possible in confidential computing into FHE-supported solutions.

\subsection{New computation environments}

The concept of confidential computing started with generic-purpose computing platforms. However, as the use of confidential computing spreads across vastly different ecosystems, the technologies need to adapt. Far edge devices have very little compute power or memory, and some of them are not even fit to run a WebAssembly runtime. TEE solutions are emerging in very small devices and microcontrollers, with a design quite different from the ones listed in this study. On the other end of the spectrum, some applications require huge amounts of computing servers, with the improvements in the High Performance Computing (HPC) ecosystem and the related design changes bringing new challenges. In these kinds of scenarios, each loss of computing power is extremely impactful. However, as the supercomputers are very costly and complex to build, they are most of the time accessible remotely and shared between users. In this situation, users may require the protection of their data, therefore turning towards confidential computing. Abstraction layers in the future may be able to tackle both ends of this spectrum by allowing the leverage of TEEs on either extremely constrained or extremely powerful systems, while ensuring that the least amount possible of computing power is lost.

Furthermore, more and more applications with specific needs require doing computations on peripherals other than the main CPU of the machine. For example, AI applications often leverage Neural Processing Units (NPUs) or Graphics Processing Units (GPUs) to speed up their computations. As the protection of a classical TEE only encompasses the main CPU, the data computed in these external peripherals is at risk of being compromised. While some efforts have already been made on extending TEEs to GPUs~\cite{wang_confidential_2024}, the ecosystem still lacks TEEs for NPUs and other accelerators, and the ones existing for GPUs are also fragmented and incompatible, resulting in the same problem as the current TEE ecosystem. An abstraction layer able to leverage a specific family of accelerators TEEs, bringing the advantages of an abstraction layer to a completely new family of applications, may be a game changer in the near future.

Another family of specialized hardware peripherals that TEE applications may need to leverage in order to do some computations are Trusted Platform Modules (TPMs) and Hardware Security Modules (HSMs). These devices are able to securely store cryptographic keys and perform cryptographic computations with them securely, without the keys leaving the hardware. TPMs and HSMs may be considered as very restricted TEEs that are only able to do very specific operations. Abstraction layers being able to securely handle the communication between various generic TEE solutions and TPMs or HSMs would allow the secure handling of custom cryptographic keys alongside the protection offered by a TEE, improving the security of the handled data. 

Some TEEs can provide additional protection for device communication, allowing the enclave to exchange data with an external hardware device securely. Such features are available in e.g. SEV-TIO~\cite{amd2023sevtio} or the IO-PMP extension of RISC-V~\cite{riscv_iopmp}. If the use of a hardware peripheral is requested by the user application, an abstraction layer should be able to detect if a TEE is able to use additional hardware peripheral protection, and if yes, correctly use this feature while keeping its use transparent for the user application. If a TEE is not compatible with such a feature, depending on the situation, the abstraction layer should be able to either use the peripheral without protection, with or without warning to the user application, or block its use if the TEE integrity is at risk. In the case of external devices, e.g., a GPU, an abstraction layer may simulate in software the presence of a GPU if a secure communication channel cannot be established, or if the currently used TEE does not support such a feature, but it will come with tremendous performance costs that will negatively impact the application.

Finally, quantum systems are a different computing environment, but they still face the same needs that led to the creation of confidential computing. Indeed, quantum computing is very complex and not likely to be generally available in the near future. Using a quantum computer most often requires the user to rent one through public offerings, making its data at risk. Quantum Blind Computing~\cite{hardware_bqc,universal_bqc} is a form of quantum computing that ensures the privacy of the computation. From this point of view, Quantum Blind Computing is close to Confidential Computing. There are several ways of achieving Quantum Blind Computing, meaning that in the future, abstraction layers for quantum blind computing may appear, bringing some of the advantages described in this study.


\section{Conclusion}
\label{sec:conclusion}

With more and more TEE solutions using various approaches and proposing different features, abstraction layers aim to unify the ecosystem, allowing application developers and system administrators to leverage confidential computing as broadly and efficiently as possible. This systematization of knowledge first listed the main TEE technologies available and described them succinctly. Emphasis has been made on summarizing this quite complex TEE ecosystem, notably because of a large amount of proprietary technologies that are not or scarcely documented. This ecosystem has also been classified in several categories depending on their main design choices. Then, abstraction layers have been listed, categorized on the main technology they rely on, and their inner workings and features are described. Finally, this study gives an extensive analysis of the ups and downs of the existing abstraction layers and on the approaches used for their implementation. One main result is that WebAssembly seems a promising approach for designing abstraction layers that are portable, secure, and support a large set of software. Future improvement possibilities have been discussed based on an analysis of the ecosystem evolution, aiming at understanding how future abstraction layers will evolve, and addressing the gaps identified in this study.\\

\noindent \textbf{Acknowledgements:} This research received funding from the Smart Networks and Services Joint Undertaking (SNS JU) under the European Union’s Horizon Europe programme: ELASTIC (GA 101139067).

\bibliographystyle{ACM-Reference-Format}
\bibliography{main.bib}

\end{document}